%% file: 00_manuscript.tex
\documentclass[a4paper,fleqn]{cas-dc}

\usepackage[numbers,sort&compress]{natbib}

\def\tsc#1{\csdef{#1}{\textsc{\lowercase{#1}}\xspace}}
\tsc{WGM}
\tsc{QE}
\tsc{EP}
\tsc{PMS}
\tsc{BEC}
\tsc{DE}

\begin{document}
\let\WriteBookmarks\relax
\def\floatpagepagefraction{1}
\def\textpagefraction{.001}
\shorttitle{}
\shortauthors{L. Malenica et~al.}

\title [mode = title]{Pore-scale mechanisms of capillary imbibition in cement-based materials revealed by direct numerical simulations in a realistic FIB-SEM microstructure}




\author[1]{Luka Malenica}[orcid=0000-0002-8309-588X]
\cormark[1]
\ead{lmalenica@ethz.ch}

\credit{Conceptualization, Methodology, Software, Investigation, Formal analysis, Data curation, Visualization, Writing – original draft, Writing -- review \& editing}

\affiliation[1]{organization={Institute for Building Materials (IfB), ETH Zurich},
                city={Zurich},
                postcode={8093}, 
                country={Switzerland}}

\author[1]{Zhidong Zhang}[]

\credit{Conceptualization, Funding acquisition, Project administration, Writing -- review \& editing}

\author[1]{Ueli Angst}[]

\credit{Conceptualization, Funding acquisition, Resources, Writing -- review \& editing}

\cortext[cor1]{Corresponding author}


\begin{abstract}
Moisture transport in cement-based materials involves transient multiphase flow through highly irregular pore networks, where evolving air--water interfaces strongly influence liquid invasion and gas displacement. Here, pore-resolved, image-based direct numerical simulations (DNS) are used to investigate spontaneous imbibition in a FIB--SEM-reconstructed cementitious microstructure, resolving air--water interface dynamics at nanometre-to-micrometre spatial scales and nanosecond-to-microsecond timescales. As these dynamics remain inaccessible to direct experimental observation at such scales, the simulations provide, to the best of our knowledge, the first direct view of transient imbibition within the capillary porosity of a realistic cementitious microstructure.

The three-dimensional simulations reveal highly heterogeneous wetting-front propagation, continuous interface reconfiguration, and progressive air entrapment, showing that local dynamics depend on evolving interface curvature and capillary-pressure fields rather than on characteristic pore size alone. Significant air entrapment develops even within the capillary pores and under perfectly water-wetting conditions, progressively reducing the saturation rate, while the capillary pressure of trapped-air clusters depends strongly on local interface geometry rather than gas volume alone. Controlled two-dimensional analyses show that the balance between capillary, viscous, inertial, and gravitational effects is strongly scale-dependent. Comparison with the three-dimensional simulations highlights the importance of out-of-plane interface curvature and pore-network connectivity. The framework also resolves moisture conditions at the steel--concrete interface, illustrating its potential for corrosion-relevant pore-scale analysis.

These results establish image-based pore-scale DNS as a virtual experiment for interrogating otherwise inaccessible fundamental moisture-transport mechanisms and provide a foundation for future extensions towards increasingly complex coupled processes relevant to cement-based-material durability.
\end{abstract}




\maketitle

\sloppy
\input{01_Introduction}

\input{01.1_Physics}

\input{02_Methodology}

\input{03_Results}

\input{04_Implications}

\input{05_Conclusions}

\printcredits

\section*{Acknowledgement}

The authors acknowledge the support from the Swiss National Science Foundation (grant number 10003479). Ueli Angst gratefully acknowledges the support of the Werner Siemens-Stiftung, the Albert Lück-Stiftung, and the Dätwyler Stiftung for the Durability of Engineering Materials research group at ETH Zurich.

\section*{Declaration of generative AI and AI-assisted technologies in the manuscript preparation process}

During the preparation of this manuscript, the authors used ChatGPT (OpenAI) to assist with language editing and refinement of the manuscript text. All AI-assisted suggestions were critically reviewed and edited by the authors, who take full responsibility for the final content of the manuscript.

\section*{Data availability}

Data will be made available on request.

\bibliographystyle{elsarticle-num-names}

\bibliography{06_Bibliography_manuscript}


\end{document}

%% file: 01_Introduction.tex
\section{Introduction}\label{sec_introduction}

Moisture transport plays a central role in the durability of cement-based materials because the amount and distribution of water within the pore structure strongly influence the transport of dissolved species, chemical reactions, freeze--thaw processes, and the conditions governing reinforcement corrosion \citep{zhang2026article,wang2024capacitive,zhou2021watersensitivity,angst2023steel,albert2026rethinking}. Among the different modes of moisture transport, capillary imbibition governs the spontaneous uptake of liquid water driven by capillary forces and is relevant to a broad range of exposure situations encountered by cementitious materials \citep{hall1989sorptivity,villagran2017improved}. Predicting how water enters and redistributes within these materials is consequently a fundamental part of understanding their transport behaviour and long-term durability.

At the material and structural scales, moisture transport is commonly described through macroscopic state variables and effective transport properties. Experimental quantities such as water uptake and degree of saturation, together with effective transport properties such as sorptivity, permeability, and moisture diffusivity, provide practical measures of the overall transport response, while continuum models describe moisture redistribution using macroscopic conservation equations and constitutive relationships formulated in terms of the material moisture state \citep{zhang2014transport,li2021water}. Depending on the formulation and conditions considered, these descriptions may represent liquid and vapour transport through different choices of state variables, constitutive relationships, and moisture-dependent transport properties \citep{celia1990general,coussy2011mechanics,zhang2016investigation,zha2019review,dong2020validation,zhang2020dual}. Such approaches are indispensable for engineering-scale prediction and provide effective descriptions of the underlying transport physics. However, continuum-scale descriptions inherently average over local fluid-solid configurations and interface dynamics, and therefore cannot resolve the pore-scale displacement and redistribution processes that collectively give rise to the observed macroscopic transport response \citep{berg2026interface}.

At the pore scale, capillary imbibition is fundamentally a transient multiphase displacement process in which wetting water advances through the pore space and displaces the non-wetting air \citep{qin2022wetting}. Classical descriptions such as the Lucas--Washburn equation provide an important physical reference by relating capillary driving pressure and viscous resistance in an idealized capillary \citep{lucasUeberZeitgesetzKapillaren1918a,washburnDynamicsCapillaryFlow1921,caiLucasWashburnEquationBasedModeling2021}. In such simple geometries, pore size and wettability constrain meniscus curvature and therefore the associated capillary pressure. Real cementitious pore structures are considerably more irregular, forming interconnected networks of constrictions, expansions, dead ends, and ink-bottle-type pores with strongly varying local dimensions and connectivity. The limitations of representing capillary transport through simple parallel-pore concepts have long been recognized for cementitious materials \citep{leventis2000capillary,moro2002inkbottle,zhang2022inkbottle}. As an air--water interface advances through such a network, its configuration continuously evolves with the local geometry, so that local curvature and capillary pressure cannot be related uniquely to a single characteristic pore size \citep{armstrong2012linking,ferrari2013direct,malenica2024towards}. Consequently, macroscopic water uptake emerges as the collective outcome of local events involving interface deformation, pinning and depinning, bypassing, reconnection, gas entrapment, and associated pressure redistribution.

Experimental methods provide increasingly detailed information on water transport across different spatial and temporal scales, ranging from integral material response to spatially resolved moisture distributions and three-dimensional fluid configurations. Conventional absorption, sorptivity, and permeability measurements quantify the integral material response \citep{hall1989sorptivity,zhou2016indirect}, whereas magnetic-resonance techniques and neutron or X-ray radiography can provide spatially resolved information on moisture redistribution and wetting-front progression over time \citep{schulteholthausen2018monitoring,zhang2011neutron,moradllo2019quantifying,zheng2022rapid}. In these measurements, however, the observed moisture state remains averaged over finite measurement volumes or, in the case of radiography, through the specimen thickness, so that the underlying pore-scale air--water interface dynamics are not resolved. Three-dimensional information can be obtained using neutron or X-ray computed tomography (CT), although the attainable spatial and temporal resolution is constrained by specimen size, acquisition time, and image quality \citep{qoku2023advances}. Previous X-ray imaging studies provided evidence of water presence and filling in larger air voids, but represented the void state in a binary manner, as either empty or filled, without resolving intermediate filling states or the evolving air--water configuration within individual voids \citep{moradllo2017using}. More recent X-ray CT studies have substantially advanced this capability by directly visualizing the moisture state and air--water configuration within macroscopic voids, including at the steel--concrete interface, and by following their evolution over prolonged exposure periods \citep{rossiXrayComputedTomography2023,mundra2025precipitation}. Most recently, the saturation evolution of more than 5000 macro-voids larger than $100~\mu$m was tracked over several months, providing a statistically rich picture of long-term void filling and trapped-air evolution \citep{governo2026revealing}. These studies provide valuable insight into long-term saturation of macroscopic voids and the secondary, slow stage of water uptake, but their spatial and temporal resolution remains insufficient to resolve the much faster primary capillary imbibition process within the finer capillary pore network.

Resolving this finer pore network requires substantially higher spatial resolution \citep{wong2022methods,jiang2022pore}. Electron-microscopy-based techniques provide access to the submicrometre pore structure, but conventional SEM is limited to two-dimensional characterization and therefore cannot fully capture pore connectivity and three-dimensional geometrical features. Focused ion beam--scanning electron microscopy (FIB--SEM) nanotomography overcomes this limitation by combining serial focused-ion-beam sectioning with SEM imaging to reconstruct the pore structure in three dimensions at nanometre-scale resolution \citep{song2019pore,song2020permeability,schmid2025probing}. This enables direct characterization of capillary porosity that is typically inaccessible to conventional X-ray or neutron tomography. However, FIB--SEM provides a static reconstruction of the material microstructure rather than the corresponding transient fluid dynamics. The resulting three-dimensional pore geometry can therefore reveal where water may move, but not how individual air--water interfaces propagate and reconfigure through the network during imbibition.

Pore-scale modelling provides a complementary means of investigating processes that remain unresolved experimentally by explicitly representing the pore geometry and fluid phases \citep{maes_geochemfoam_2021}. The level of representation can vary considerably \citep{golparvar2018comprehensive,zhao2019comprehensive}. Simplified approaches such as capillary-bundle and pore-network models describe the pore space through idealized or reduced geometrical elements and simplified flow equations, making them computationally efficient \citep{dong2005immiscible,xiong2016review,blunt_multiphase_2017}. At a higher level of geometrical and physical resolution, pore-resolved simulations solve the fluid motion directly within the actual pore space, allowing local velocities, pressures, phase distributions, interface configurations, and connectivity to emerge from the imposed geometry and governing physics. Such simulations are commonly performed using lattice Boltzmann methods (LBM) or direct numerical approaches based on the Navier--Stokes equations \citep{zhao2019comprehensive,chen2022pore}. 

Outside cement-based materials, image-based pore-resolved modelling is well established in geoscience and digital-rock physics, where three-dimensional pore geometries reconstructed predominantly by X-ray microtomography are used to investigate transient multiphase displacement, phase trapping, connectivity, and relative permeability \citep{blunt2013pore,wildenschild2013xray,raeini2014direct,shams2021direct,santos2022mplbm}. Extending such approaches to cementitious materials is considerably more demanding because a substantial part of the transport-relevant pore network extends into the submicrometre range, placing much stronger requirements on both microstructural imaging and computation.

Within cementitious materials, these challenges have favoured the use of simplified capillary-bundle and pore-network approaches to relate pore-size distributions, connectivity, and saturation state to effective transport properties such as permeability and diffusivity \citep{nguyen2010capillary,zhang2019water,shi2023modified,babaei2021permeability}. At intermediate scales, two-phase water transport in concrete has also been addressed using mesoscopic approaches, including coupled DEM--CFD formulations in which liquid and gas transport is represented through reduced flow networks \citep{krzaczek2023modelling}. Tong et al. \citep{tong2025advanced} recently used a simplified pore-scale approach on numerically generated cement-paste microstructures to represent desorption, assuming that pore emptying depends on whether the local pore size exceeds a critical pore diameter. The more computationally demanding pore-resolved studies remain comparatively limited and have relied predominantly on LBMs. Zalzale and McDonald \citep{zalzale2012lattice} used a free-energy multiphase LBM to simulate liquid--vapour distributions and capillary adsorption--desorption in numerically generated cement-paste microstructures, with the fluid phases re-equilibrated as vapour pressure was varied. Zalzale et al. \citep{zalzale2013three} subsequently examined water and gas permeability at partial saturation, with phase distributions prescribed according to the Kelvin--Laplace relation. Related studies adopted a multiphase strategy in which the degree of water saturation was prescribed and the liquid and gas phases were typically initialized randomly before being allowed to redistribute towards equilibrium or stable configurations. The resulting phase configurations were used to investigate ionic diffusivity, including chloride transport, and oxygen diffusivity \citep{zhang2012ionic,liu2022ionic,liu_multiscale_2022,liu2023chloride,liu2023oxygen}, phase connectivity and critical saturation \citep{zhang2014moisture,zhang2017pore}, and water and gas permeability \citep{zhang2017pore}. More recently, Li et al. \citep{li2024study} examined how saturation degree, hydration age, wettability, and pore structure affect the resulting water distribution and permeability. Overall, the multiphase evolution in these studies primarily served to establish equilibrium or stable phase configurations for subsequent transport analysis, rather than to investigate the transient dynamics of water ingress itself.

An additional important consideration is the origin and spatial scale of the pore geometries used in these studies. Most previous LBM investigations relied on numerically generated cement-paste microstructures obtained from hydration models and subsequently discretized for computation \citep{zalzale2012lattice,zalzale2013three,zhang2012ionic,liu2022ionic,liu2023chloride,liu2023oxygen,li2024study}. The only multiphase studies using experimentally reconstructed pore geometries identified here are those of Zhang and co-workers, who used X-ray micro-CT geometries with a voxel size of approximately \(0.5~\mu\mathrm{m}\) \citep{zhang2014moisture,zhang2017pore}. At this voxel size, sub-micrometre pores are not reliably resolved, leaving a substantial part of the finer sub-micrometre capillary pore network outside the explicitly represented geometry. Consequently, local pore-scale features relevant to meniscus deformation, invasion pathways, and air entrapment during water ingress remain substantially unresolved.

Building on our earlier DNS studies using simplified two-dimensional pore geometries and idealized steel--concrete interfacial voids \citep{malenica2024towards,mundra2025precipitation}, the present work extends this framework to a substantially more complex image-based three-dimensional cementitious pore network. The study combines high-resolution FIB--SEM nanotomography of the pore structure reconstructed by Schmid et al. \citep{schmid2025probing} with Navier--Stokes-based pore-resolved direct numerical simulations. Only a few studies have so far combined high-resolution FIB--SEM imaging with pore-resolved multiphase flow simulations, with applications reported for gas-diffusion electrodes, thin coating layers, shale nanopores, and lithium-ion battery electrodes \citep{kunz2020prediction,aslannejad2021liquid,qin2024three,pazdniakou2026investigation}; however, no such application has been reported for cement-based materials. Applied here to a realistic cementitious pore structure, this framework provides direct access to the transient invasion of water into an initially dry pore network, resolving the evolving air–water interfaces, local pressure fields, interface curvature, phase connectivity, and air entrapment. At a voxel size of $29$ nm and over nanosecond-to-microsecond timescales, the simulations capture pore-scale dynamics that remain inaccessible to direct experimental observation at comparable spatial and temporal resolution. To the best of our knowledge, this represents the first pore-resolved numerical description of transient imbibition within a realistic cementitious pore structure at this combination of spatial and temporal scales.

The remainder of the paper is organized as follows. Section~\ref{subsec_physics_SI} introduces the physical principles of capillary imbibition through analytical descriptions in simplified geometries, providing a reference framework for interpreting the more complex pore-scale simulations and for identifying where such idealized descriptions remain applicable or break down. Section~\ref{sec_methodology} describes the computational geometry, material and fluid properties, and the numerical methodology used for the pore-resolved simulations. Section~\ref{sec_results} presents the numerical results, including spontaneous imbibition in the realistic three-dimensional FIB--SEM pore structure, complementary analyses using simplified and systematically rescaled two-dimensional geometries to isolate individual mechanisms, and their interpretation within the full three-dimensional pore network, together with an application to moisture conditions at the steel--concrete interface. Section~\ref{sec_discussion} discusses the broader physical implications, model scope, and opportunities for pore-scale DNS in cementitious materials, before the main conclusions are summarized in Section~\ref{sec_conclusions}.

%% file: 01.1_Physics.tex
\section{Physical principles of capillary imbibition}
\label{subsec_physics_SI}

\begin{figure*}[!t]
\centering
\includegraphics[width=\textwidth]{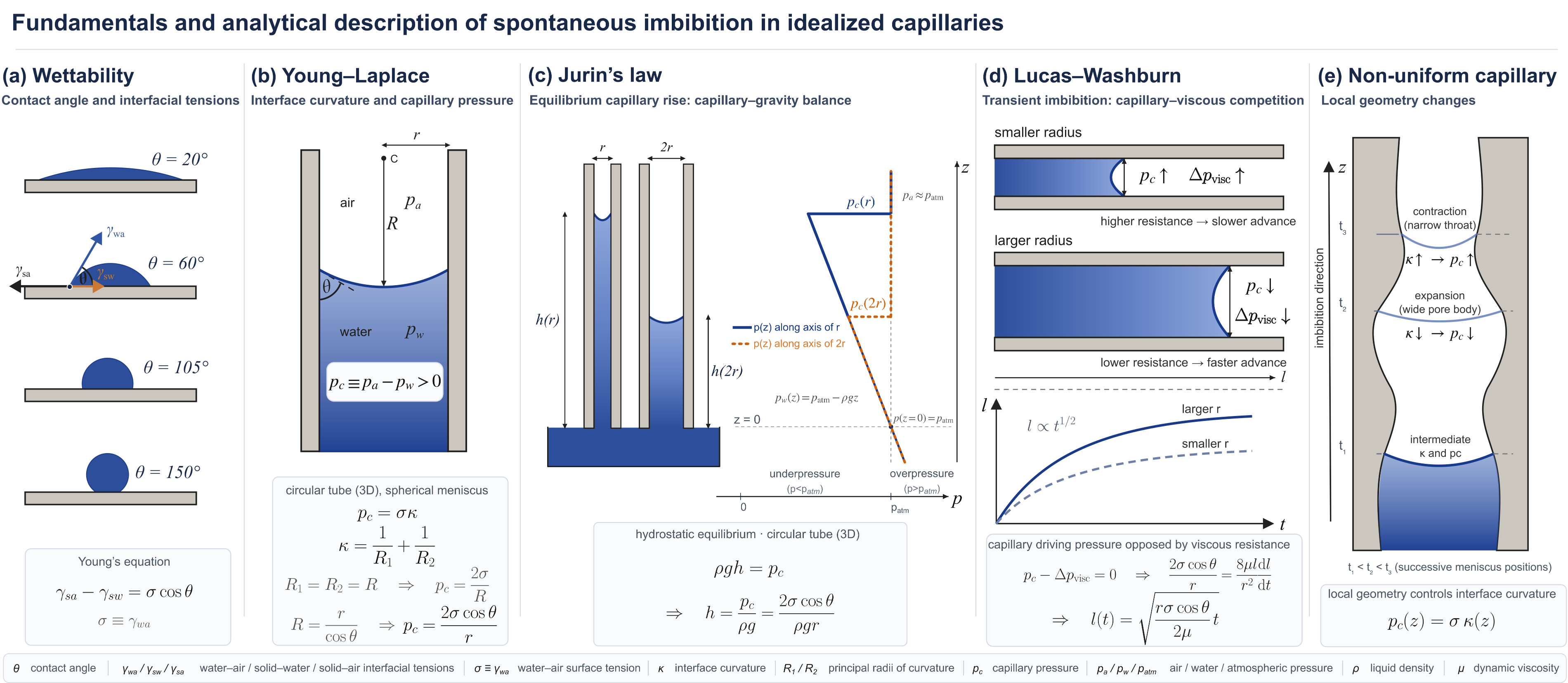}
\caption{Fundamental physical principles governing spontaneous imbibition in idealized capillary geometries: (a) wettability and equilibrium contact angle (Young's equation), (b) capillary pressure arising from interface curvature (Young--Laplace equation), (c) equilibrium capillary rise (Jurin's law), (d) transient imbibition in an isolated capillary (Lucas--Washburn equation), (e) effect of non-uniform capillary geometry on interface curvature and capillary pressure.}
\label{fig_fundamentals}
\end{figure*}

Spontaneous imbibition in porous materials is governed by the interplay between capillary, viscous, inertial, and gravitational forces, together with wettability and pore geometry. In cementitious materials, these mechanisms operate within highly irregular and interconnected three-dimensional pore networks, making spontaneous imbibition considerably more complex than can be represented by classical analytical models derived for simplified capillary geometries. Nevertheless, despite their restrictive physical assumptions, these analytical models provide fundamental physical insight into the governing mechanisms of spontaneous imbibition and serve as an accessible introduction to the underlying physics before addressing the more complex behaviour occurring in realistic cementitious microstructures. Figure~\ref{fig_fundamentals} summarizes the fundamental concepts discussed in this section.

Capillary imbibition originates from the combined action of surface tension and wettability (Figure~\ref{fig_fundamentals}a). Surface tension arises from the imbalance of cohesive molecular forces at the fluid--fluid interface and continuously drives the interface towards configurations of minimum interfacial energy. Consequently, isolated droplets and bubbles tend to adopt spherical shapes corresponding to the minimum surface area for a given volume. When the fluid--fluid interface comes into contact with a solid surface, its shape is additionally influenced by the wettability of that surface. Wettability describes the affinity of a liquid for a solid surface in the presence of another immiscible fluid and is commonly characterized by the equilibrium contact angle (Figure~\ref{fig_fundamentals}a), which follows from Young's equation,

\begin{equation}
\gamma_{sa}-\gamma_{sw}=\sigma\cos\theta,
\end{equation}

where $\gamma_{sa}$, $\gamma_{sw}$, and $\sigma$ denote the solid--air, solid--water, and water--air interfacial tensions, respectively, and $\theta$ is the equilibrium contact angle measured through the wetting phase. Although wettability is fundamentally determined by the balance of these interfacial tensions, it is typically characterized experimentally by the equilibrium contact angle, which serves as the primary descriptor of surface wettability. Consequently, surfaces with $\theta<90^\circ$ are classified as hydrophilic and promote spontaneous liquid invasion, whereas surfaces with $\theta>90^\circ$ are classified as hydrophobic and oppose it.

Within porous media, the local wettability (contact angle) together with the surrounding solid boundaries determine the shape of the liquid--gas interface, giving rise to curved menisci within the pore space. The resulting interface curvature, combined with the surface tension, generates a pressure difference between the wetting and non-wetting phases, referred to as the capillary pressure,

\begin{equation}
p_c=p_a-p_w,
\label{eq:pc}
\end{equation}

where $p_a$ and $p_w$ denote the pressures in the air (non-wetting) and water (wetting) phases, respectively. The Young--Laplace equation relates this pressure difference to the local interface curvature (Figure~\ref{fig_fundamentals}b),

\begin{equation}
p_c=\sigma\kappa,
\label{eq:yle}
\end{equation}

where \(\kappa=\kappa_1+\kappa_2\) is the total interface curvature, with \(\kappa_1\) and \(\kappa_2\) representing the principal curvatures along two orthogonal tangent directions at the air--water interface. An important consequence of the Young--Laplace equation is that capillary pressure is fundamentally governed by the local interface curvature rather than by pore size itself. Only for idealized capillary geometries, in which the interface curvature is uniquely determined by the local geometry (Figure~\ref{fig_fundamentals}b), can the capillary pressure be expressed directly as a function of the capillary dimensions. For cylindrical capillary of radius $r$, this reduces to

\begin{equation}
p_c=\frac{2\sigma\cos\theta}{r}.
\end{equation}

This expression shows that smaller capillaries generate larger capillary pressures owing to their greater interface curvature.

The simplest application of this relationship is the equilibrium capillary rise in a vertical cylindrical capillary. As the liquid column rises, the hydrostatic pressure difference across the liquid column (\(\rho g h\)) increases until it balances the capillary pressure (Figure~\ref{fig_fundamentals}c), yielding Jurin's law

\begin{equation}
h=\frac{2\sigma\cos\theta}{\rho g r},
\label{eq:jurin}
\end{equation}

where $h$ is the equilibrium capillary rise height, $\rho$ is the liquid density, and $g$ is the gravitational acceleration.

Jurin's law describes only the final equilibrium state after the liquid has ceased to move. It provides no information about the transient evolution of the advancing interface. Before equilibrium is reached, liquid motion results from the combined effects of capillary pressure, viscous resistance, inertia, and gravity.

The transient imbibition process is commonly described by the Lucas--Washburn equation. Under the simplifying assumptions that inertial and gravitational effects are negligible, balancing the capillary driving pressure with the viscous pressure loss (Figure~\ref{fig_fundamentals}d) yields

\begin{equation}
l(t)=\sqrt{\frac{r\sigma\cos\theta}{2\mu}t},
\label{eq:lwe}
\end{equation}

where $l$ is the penetration length, $t$ is time, and $\mu$ is the dynamic viscosity of the liquid.

An important consequence of this solution is that larger isolated capillaries exhibit faster transient imbibition despite their lower capillary pressure. This apparently counterintuitive behaviour results from the different scaling of the capillary driving pressure and viscous resistance. While the capillary pressure scales as $p_c\propto r^{-1}$, the viscous pressure loss scales as $\Delta p_\mu\propto r^{-2}$. Consequently, decreasing the capillary radius increases viscous resistance more rapidly than capillary pressure. Narrower capillaries therefore exhibit slower transient advancement despite their greater capillary driving pressure.

The Lucas--Washburn equation (~\ref{eq:lwe}) is derived for an isolated capillary of constant cross-section, in which the interface curvature, and therefore the capillary pressure, are uniquely determined by the local capillary geometry. Figure~\ref{fig_fundamentals}e illustrates how this relationship changes even within a single capillary of non-uniform geometry. As the interface propagates through pore constrictions and widenings, it continuously changes shape, causing both the interface curvature and the corresponding capillary pressure to evolve along the flow path.

Furthermore, real cementitious materials introduce an additional level of complexity because their pores form highly irregular and strongly interconnected three-dimensional networks. Consequently, spontaneous imbibition in interconnected pore networks is considerably more complex, often resulting in fundamentally different imbibition behaviour than predicted by the Lucas--Washburn equation for isolated capillaries \citep{ashraf2019capillary,malenica2024towards}. Thus, while the analytical models discussed above provide valuable physical insight into the governing mechanisms of spontaneous imbibition, they cannot fully capture the evolution of interface curvature, capillary pressure, and liquid redistribution within realistic cementitious microstructures. 

Unlike the simplified analytical models, the present work solves the full Navier--Stokes equations within realistic cementitious microstructures rather than idealized capillary geometries. All relevant pressure, viscous, inertial, gravitational, and surface-tension effects are retained, while wettability is prescribed through the contact angle at the three-phase (air--water--solid) contact line. Consequently, interface curvature, capillary pressure, and liquid redistribution emerge naturally from the governing equations and the resolved pore geometry, providing direct physical insight into pore-scale mechanisms that cannot be captured by simplified analytical models.

%% file: 02_Methodology.tex
\section{Methodology}\label{sec_methodology}
\subsection{Numerical model}
\label{subsec_fullEQ}

\subsubsection{Navier--Stokes equations}
\label{subsec_NS}

To resolve spontaneous imbibition directly at the pore scale, the incompressible flow of two immiscible Newtonian fluids is described using a one-fluid formulation of the Navier--Stokes equations (NSE). The governing equations for mass and momentum conservation are given by \citep{tryggvason_direct_2011}:

\begin{equation}
\nabla \cdot \mathbf{u}=0
\label{eq:MassConserv}
\end{equation}

\begin{equation}
\frac{\partial \rho \mathbf{u}}{\partial t}
+\nabla \cdot (\rho \mathbf{u}\mathbf{u})
=
-\nabla p
+\nabla \cdot
\left\{
\mu
\left[
\nabla \mathbf{u}
+
(\nabla \mathbf{u})^T
\right]
\right\}
+\rho\mathbf{g}
+\mathbf{f}_{st}
\label{eq:MomentumConserv}
\end{equation}

where $\mathbf{u}$ is the velocity vector, $t$ is time, $p$ is pressure, $\rho$ is density, $\mu$ is dynamic viscosity, $\mathbf{g}$ is the gravitational acceleration, and $\mathbf{f}_{st}$ is the surface-tension body force.

For constant interfacial tension, the surface-tension force is represented using the Continuous Surface Force (CSF) formulation proposed by \citet{brackbill_continuum_1992},

\begin{equation}
\mathbf{f}_{st}
=
\sigma
\kappa
\mathbf{n}
\delta_{\Gamma}
\label{eq:CSF}
\end{equation}

where $\sigma$ is the interfacial tension, $\kappa$ is the interface curvature, $\mathbf{n}$ is the unit normal vector to the fluid--fluid interface, and $\delta_\Gamma$ is the Dirac delta function localized at the interface.

Equations~(\ref{eq:MassConserv})--(\ref{eq:MomentumConserv}) are valid throughout the computational domain, even when fluid properties such as density and viscosity change discontinuously across the fluid interface. To solve these equations, the location of each fluid within the domain must therefore be identified. In this work, the Volume of Fluid (VOF) interface-capturing approach \citep{maric2020unstructured,kim2021efficient} is used to distinguish between the two fluids on a fixed computational grid.

\subsubsection{Volume of Fluid (VOF) method}
\label{subsec_VOF}

The algebraic Volume of Fluid (VOF) interface-capturing approach \citep{hirt_volume_1981} is used to track the evolution of the air--water interface on a fixed computational grid. In the VOF method, the interface is represented by an indicator (volume fraction) function $\alpha$, where $\alpha=1$ denotes the first fluid, $\alpha=0$ the second fluid, and the fluid--fluid interface is represented by the narrow region over which \(\alpha\) changes rapidly between 0 and 1.

Using the indicator function, the fluid density and dynamic viscosity are interpolated as

\begin{equation}
    \rho=\alpha \rho_1 + (1-\alpha)\rho_2
    \label{eq:vof_rho}
\end{equation}

\begin{equation}
    \mu=\alpha \mu_1 + (1-\alpha)\mu_2
    \label{eq:vof_mu}
\end{equation}

where $\rho_i$ and $\mu_i$ denote the density and dynamic viscosity of phase $i$, respectively.

The evolution of the interface is obtained by solving the transport equation for the indicator function,

\begin{equation}
    \frac{\partial \alpha}{\partial t}
    +
    \nabla\cdot(\alpha \mathbf{u})
    =
    0
    \label{eq:VOF}
\end{equation}

In the numerical implementation, interface compression is often used to limit numerical diffusion and maintain a sharp fluid--fluid interface. The interface normal vector required by the CSF formulation (Eq.~\ref{eq:CSF}) is evaluated from the indicator function as

\begin{equation}
    \mathbf{n}
    =
    \frac{\nabla\alpha}
    {||\nabla\alpha||}
    \label{eq:vof_normal}
\end{equation}

while the interface curvature is obtained from

\begin{equation}
    \kappa
    =
    \nabla\cdot\mathbf{n}
    =
    \nabla\cdot
    \left(
    \frac{\nabla\alpha}
    {||\nabla\alpha||}
    \right)
    \label{eq:vof_curvature}
\end{equation}

To prescribe wettability at solid boundaries, the interface normal vector used for curvature evaluation is adjusted near the solid surface to enforce the prescribed contact angle,

\begin{equation}
    \mathbf{n}
    =
    \mathbf{n}_s\cos\theta
    +
    \mathbf{t}_s\sin\theta
    \label{eq:normal_BC}
\end{equation}

where $\mathbf{n}_s$ and $\mathbf{t}_s$ are the unit normal and tangent vectors to the solid surface, respectively.

In this work, the governing equations are solved using the \textit{interGCFoam} solver implemented in the GeoChemFoam framework. GeoChemFoam is an OpenFOAM-based finite-volume code developed for direct numerical simulations of pore-scale transport processes in realistic porous media and has been successfully applied to a wide range of single-phase and multiphase, reactive and non-reactive transport problems \citep{maes_direct_2018,maes_new_2018,maes_unified_2020,maes_improved_2022,
menke2023channeling,maes2024dispersivity,yong2024multiscale,
yang2023coupledCO2,abdellatif2026benchmark,liu2026nanoconfined,
liu2026nanomicrocontinuum}. More numerical and implementation details are provided in \citet{maes_geochemfoam_2021,maes2021geochemfoam}.

\subsection{Computational geometry and material properties}

\subsubsection{Image-based digital twin of reinforced mortar}

\begin{figure*}
\centering{\includegraphics[width=\textwidth]{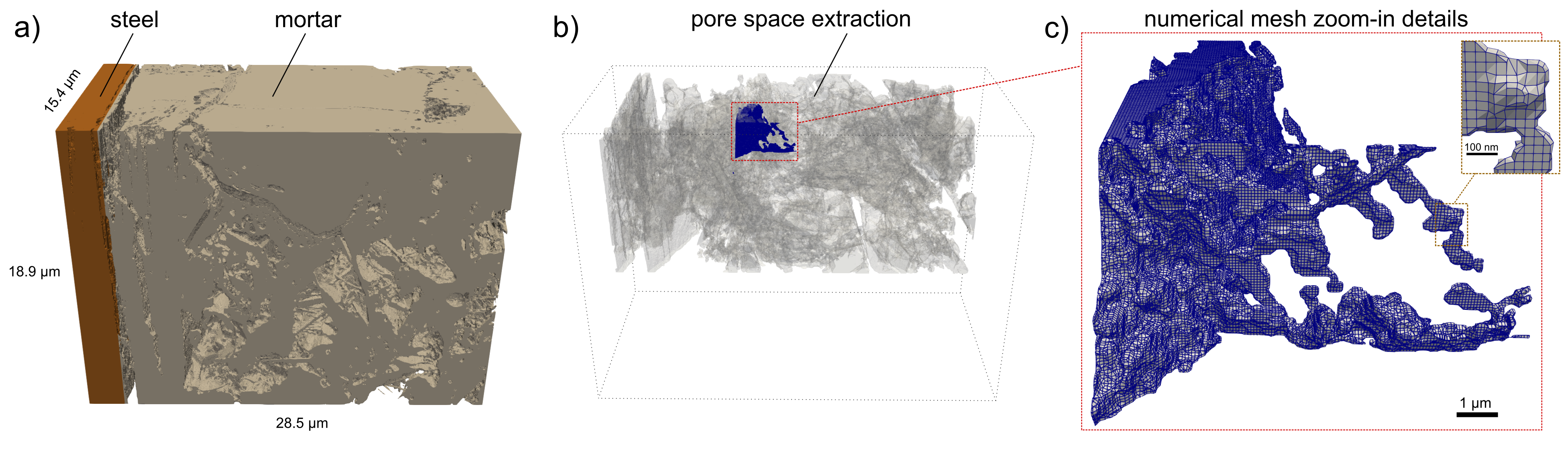}}
\caption{3D FIB-SEM nanotomography: a) Visualization of the complete reconstructed reinforced mortar microstructure (solid material), b) extracted pore (void) space from the sample used for the 3D simulations, c) close-up view of a small portion of the finite-volume mesh used to discretize the pore space. The extracted pore-space domain (b) comprises \(820\times335\times500\) (~137 million) voxels at 29 nm resolution, corresponding to physical dimensions of approximately \(23.8\times9.7\times14.5~\mu\mathrm{m}^3\). A three-dimensional visualization of the mortar microstructure together with a fly-through of the extracted pore space is provided in the Supplementary Material (anim-01).}
\label{fig_geometry}
\end{figure*}

Direct numerical simulations require a three-dimensional representation of the pore space with sufficient spatial resolution to resolve the capillary pathways governing air--water displacement. In the present work, the FIB--SEM nanotomography dataset reconstructed by \citet{schmid2025probing} is used as the computational geometry. The dataset provides a high-resolution three-dimensional digital twin of the reinforced mortar microstructure at a non-corroded steel--concrete interface (Fig.~\ref{fig_geometry}a), with a voxel of 29 nm. The complete reconstructed volume consists of 969 $\times$ 711 $\times$ 514 voxels (approximately 354 million voxels).

The reconstructed microstructure has a porosity of approximately 17\%, indicating that the majority of the volume is occupied by the solid skeleton. Nevertheless, direct numerical simulation of the complete pore-space domain would still result in a prohibitively large computational problem. Therefore, a sub-volume of 820 $\times$ 335 $\times$ 500 voxels (approximately 137 million voxels) was extracted for the present study (Figure~\ref{fig_geometry}b). To accurately resolve both the complex pore geometry and the evolving air--water interface, the extracted pore space was discretized into approximately 38 million finite-volume cells (Figure~\ref{fig_geometry}c).

\subsubsection{Fluid properties, wettability and simulation setup}

The pore space was assumed to contain air and water as two incompressible, immiscible Newtonian fluids. The densities and dynamic viscosities of water and air were set to $\rho_w=1000$~kg\,m$^{-3}$, $\rho_a=1.225$~kg\,m$^{-3}$, $\mu_w=1.0\times10^{-3}$~Pa\,s, and $\mu_a=1.81\times10^{-5}$~Pa\,s, respectively. The air--water surface tension was set to $\sigma=0.072$~N\,m$^{-1}$, and gravitational acceleration $g=9.81$~m\,s$^{-2}$ was applied in the vertical (negative \(y\)) direction.

The reconstructed solid skeleton was assumed to be perfectly water-wet ($\theta = 0^\circ$), reflecting the strongly water-wetting nature of cementitious materials. This assumption provides the maximum capillary driving force and allows the fundamental pore-scale mechanisms governing spontaneous imbibition to be investigated without the additional complexity associated with heterogeneous wettability. The influence of heterogeneous wettability is beyond the scope of the present work and will be investigated in future studies.

The spontaneous imbibition problem is defined by the following initial and boundary conditions. Initially, the pore space above the prescribed water level is filled with air, with a flat air–water interface separating the two phases. The top and bottom boundaries are defined as free inlet and outlet boundaries, allowing both air and water to enter or leave the computational domain. Atmospheric pressure, corresponding to zero gauge pressure (\(p=0\)), is prescribed at the top boundary, while the initial water level represents the free surface of a connected external water reservoir in an equivalent experimental configuration. The remaining boundaries correspond to the reconstructed solid skeleton and are treated as impermeable no-slip walls with a prescribed equilibrium contact angle of \(\theta = 0^\circ\).

%% file: 03_Results.tex
\section{Results}
\label{sec_results}


\subsection{Pore-scale insight using an image-based 3D cementitious pore space}
\label{subsec_3D_SI}

\subsubsection{Spatio-temporal evolution of wetting front}
\label{subsubsec_3D_wetting_saturation}

\begin{figure*}
\centering{\includegraphics[width=\textwidth]{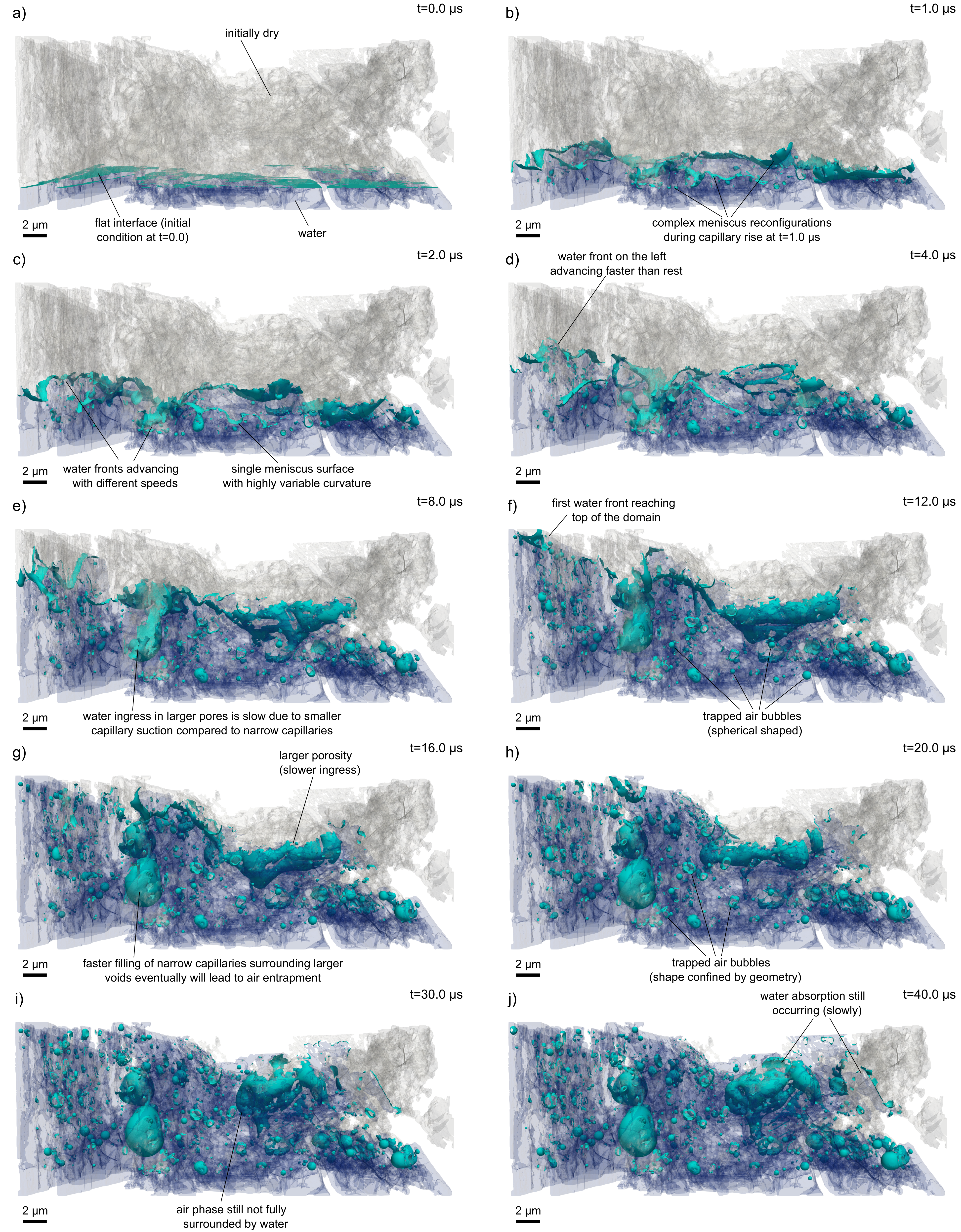}}
\caption{Numerical simulation of air--water multiphase flow during spontaneous capillary imbibition through the realistic FIB--SEM pore space, showing highly heterogeneous water ingress through the irregular pore network, even within micrometre-scale reconstructed volume. Water and air are shown in transparent blue and white, respectively, and the evolving air--water interfaces by green surfaces. The complete animation is provided in the Supplementary Material (anim-02).}\label{fig_3Dfull}
\end{figure*}

\begin{figure}[t]
\centering
\includegraphics[width=0.98\columnwidth]{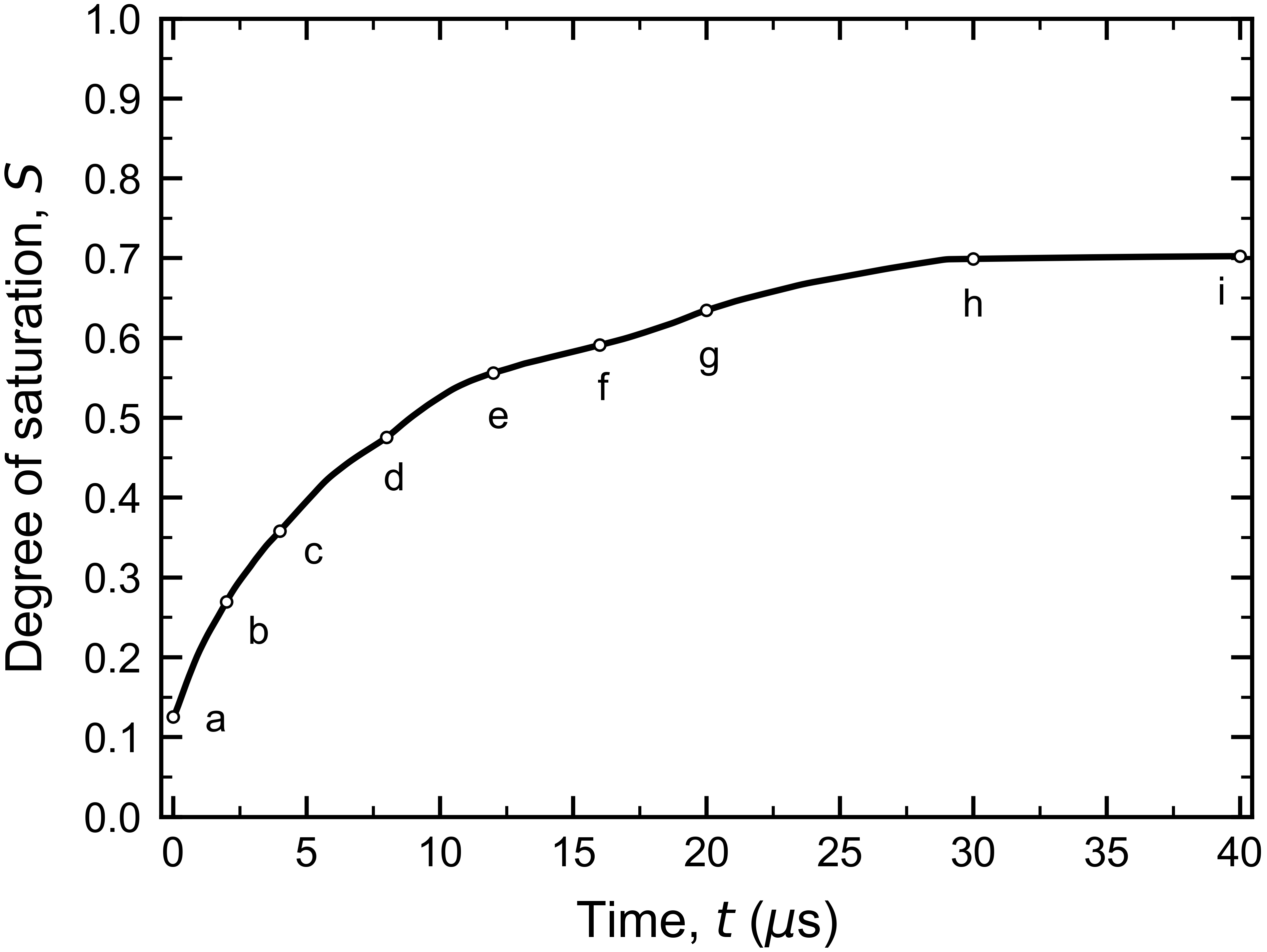}
\caption{Evolution of the degree of saturation during spontaneous imbibition in the reconstructed three-dimensional pore network. Open circles indicate the time instants corresponding to the snapshots shown in Figure~\ref{fig_3Dfull}.}
\label{fig_S_t}
\end{figure}

Figure~\ref{fig_3Dfull} presents the simulated spontaneous capillary rise through the FIB--SEM pore space. The water and air phases occupying the pore space are shown in transparent blue and white, respectively, while the air--water interfaces are highlighted by green surfaces. The cementitious solid skeleton is omitted for clarity. The corresponding degree of saturation is shown in Figure~\ref{fig_S_t}.

Starting from the initially flat air–water interface shown in Figure~\ref{fig_3Dfull}a, and in the absence of any externally imposed pressure difference, the combination of surface tension and the strongly water-wetting nature of the cementitious material causes the interface to rapidly evolve into a highly irregular collection of air–water menisci (Figure~\ref{fig_3Dfull}b). The resulting capillary pressure drives spontaneous water rise against gravity. In fact, the hydrostatic pressure difference across the approximately \(10~\mu\mathrm{m}\) vertical extent of the domain is only \(0.1~\mathrm{Pa}\), many orders of magnitude smaller than the characteristic capillary pressures at this scale. Thus, although gravity is included in the simulations, its influence on imbibition can be considered negligible at the original FIB--SEM scale (its increasing importance at larger scales is investigated in Section~\ref{subsec_size_effects}). As spontaneous imbibition progresses, the number, shape, and spatial distribution of the air--water menisci continuously evolve, producing an increasingly complex wetting front, as shown in the remaining snapshots of Figure~\ref{fig_3Dfull}.

As discussed in Section~\ref{subsec_physics_SI}, the capillary pressure is determined by the surface tension and the local interface curvature according to the Young--Laplace equation (Eq.~\ref{eq:yle}). With surface tension prescribed as constant and wettability represented by a uniform equilibrium contact angle, the pore-space geometry becomes the primary factor controlling the interface curvature and, consequently, the local capillary pressure.

This behaviour is clearly illustrated in Figure~\ref{fig_3Dfull}c, where even a single connected air--water interface exhibits substantial spatial variations in curvature, resulting in pronounced local variations in capillary pressure along the same interface. Consequently, different parts of the same connected interface advance at different rates, while neighbouring pores may exhibit markedly different invasion dynamics despite experiencing identical fluid properties and boundary conditions.

One consequence of this geometry-controlled heterogeneity is preferential flow through small-aperture features, as observed in a thin crack near the left edge of the domain (Figure~\ref{fig_3Dfull}d). Owing to its small aperture, the crack generates a highly curved interface and consequently a high capillary pressure, causing the wetting front to propagate rapidly towards the top of the domain (Figures~\ref{fig_3Dfull}b--f).

A contrasting consequence is the delayed filling of larger pores, as observed in the right part of the domain (Figure~\ref{fig_3Dfull}g). Unlike the classical Lucas--Washburn model (Figure~\ref{fig_fundamentals}d), where imbibition in each capillary is considered independently, spontaneous imbibition within an interconnected pore network is governed by local pressure gradients established throughout the connected liquid phase \citep{malenica2024towards}. As the interface advances into a larger pore, its curvature generally decreases, reducing the local capillary pressure (Figure~\ref{fig_fundamentals}e), whereas the more strongly curved menisci in the surrounding finer pores generate greater capillary suction. Because these interfaces are hydraulically connected through the liquid phase, the resulting differences in capillary pressure drive preferential water invasion through the finer pore network, despite its greater viscous resistance. Consequently, the larger pore fills more slowly while the surrounding finer pores continue to advance. This behaviour is particularly evident in Figures~\ref{fig_3Dfull}f--i, where the wetting front reaches the upper left part of the domain after approximately 12~$\mu$s, whereas water continues to advance locally around the larger pore even at 40~$\mu$s. This behaviour is consistent with the pore-scale mechanisms reported by \citet{ashraf2023spontaneous} and \citet{malenica2024towards}.

\subsubsection{Air entrapment mechanisms and degree of saturation}
\label{subsubsec_air_entrapment}

As spontaneous imbibition progresses, portions of the gas phase become progressively isolated, leading to air entrapment within the pore network. Depending on their interaction with the surrounding water phase and the solid skeleton, two characteristic trapped-air configurations can be distinguished.

The first configuration consists of almost perfectly spherical trapped-air bubbles (marked in Figure~\ref{fig_3Dfull}f). Such shapes arise because surface tension minimizes the interfacial energy by minimizing the air--water interfacial area, for which a sphere represents the minimum-area configuration for a given volume. Accordingly, trapped-air bubbles detached from the solid surface and completely surrounded by the water phase are expected to adopt such a nearly spherical equilibrium shape.

The second configuration consists of trapped-air volumes that remain connected to the solid skeleton, with part of their boundary formed by the air--water interface and the remainder by the surrounding solid (marked in Figure~\ref{fig_3Dfull}h). In some cases, these represent attached air bubbles that do not detach from the solid surface despite the prescribed zero contact angle, likely promoted by the combined effects of transient interface dynamics and the highly irregular pore geometry, including contact-line pinning induced by the geometrically resolved surface roughness \citep{mehmani2019capillary}. In real cementitious materials, additional contact-angle hysteresis and spatial variations in surface wettability are expected to further complicate contact-line dynamics and may therefore further promote the persistence of such attached air bubbles. 
In other cases, trapped air remains confined within dead-end pores after the advancing meniscus spans and seals the pore entrance, disconnecting the enclosed air from the external gas phase. As a result, the remaining air volume is retained within the dead-end pore and remains bounded directly by the surrounding solid skeleton. Unlike the attached bubbles discussed above, these trapped-air volumes arise directly from the local pore topology rather than from incomplete bubble detachment from the solid surface \citep{yu2024unraveling}.

In both cases associated with this second trapped-air configuration, the rendered air--water interface appears to contain openings (Figure~\ref{fig_3Dfull}h). This is because the visualization shows only the portion of the trapped-air boundary formed by the air--water interface, whereas the remaining boundary is defined by the surrounding solid skeleton and not shown. Following entrapment, the pressure within each isolated trapped-air volume tends towards a uniform value. The air--water interface therefore tends towards a constant-mean-curvature configuration, consistent with the nearly uniform pressure jump across the interface. Because only part of the trapped-air boundary is deformable, this relaxation can occur only through reconfiguration of the fluid--fluid interface, potentially accompanied by displacement of the contact line, whereas the solid-boundary portion remains defined by the local pore geometry.

\begin{figure*}
\centering{\includegraphics[width=\textwidth]{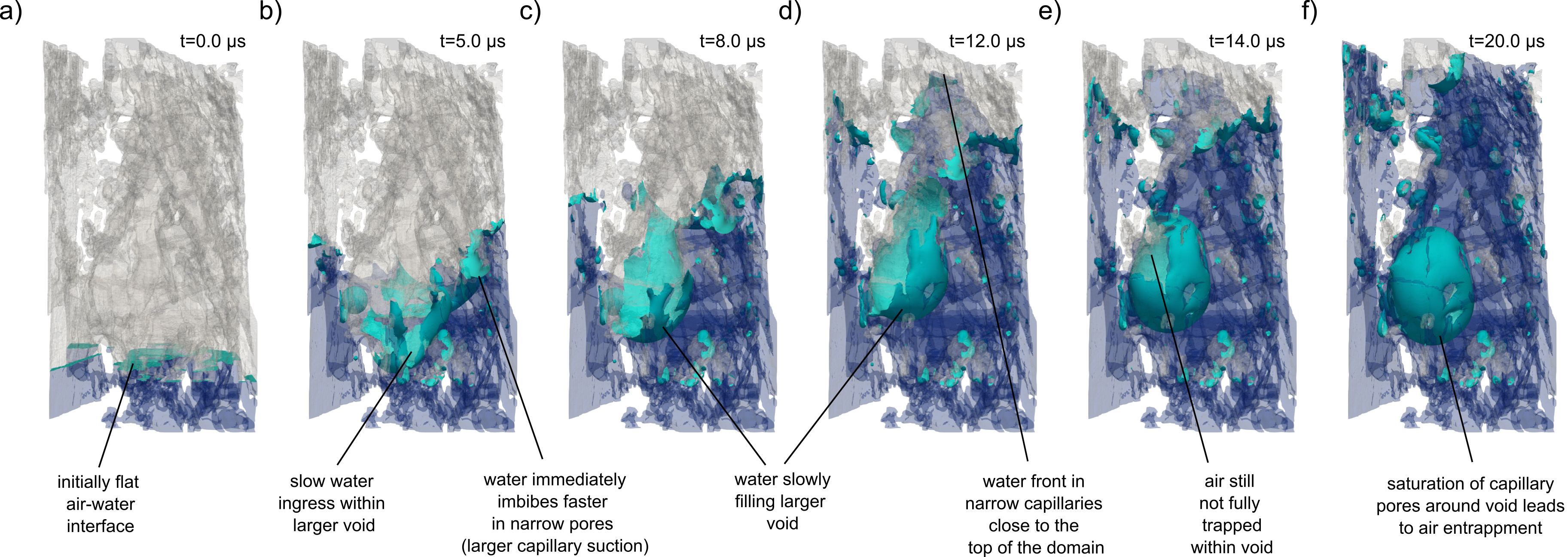}}
\caption{Insight into air entrapment within a larger pore. Within the interconnected pore network, the surrounding finer pores develop stronger capillary suction and fill more rapidly than the larger pore, despite their greater viscous resistance. The neighbouring wetting fronts therefore advance around the more slowly filling pore and eventually reconnect, disconnecting the remaining air from the external gas phase and producing residual air entrapment. The complete animation is provided in the Supplementary Materials (anim-02c).}\label{fig_void_filling}
\end{figure*}

Among the pore-scale processes contributing to air entrapment, the delayed filling of larger pores described above is examined here in more detail. Figure~\ref{fig_void_filling} presents a magnified view of a representative large pore and illustrates how the different filling rates of the large pore and the surrounding finer pore network ultimately result in residual air entrapment.

Figure~\ref{fig_void_filling}a shows the initial configuration before the onset of spontaneous imbibition. As spontaneous imbibition begins, water simultaneously invades the large pore and the surrounding finer pore network (Figures~\ref{fig_void_filling}b,c). As established in previous section, the surrounding finer pore network fills more rapidly than the large pore, so that the neighbouring wetting fronts progressively advance around the more slowly filling pore. As spontaneous imbibition continues, the large pore becomes progressively surrounded by the invading water phase, while its remaining connection to the external gas phase becomes increasingly restricted (Figures~\ref{fig_void_filling}c,d). Eventually, the surrounding wetting fronts reconnect around the large pore (Figures~\ref{fig_void_filling}e,f), sealing its remaining connection to the external gas phase and thereby trapping the enclosed air. Following disconnection, the entrapped air gradually relaxes under the action of surface tension towards a more compact and energetically favourable configuration, with the air--water interface reconfiguring within the constraints imposed by the surrounding pore geometry and the contact line along the solid surface.

This sequence provides direct pore-scale evidence for the conceptual mechanism proposed by \citet{fagerlund2004service}. By resolving the transient three-dimensional evolution of differential water invasion and the subsequent reconnection of wetting fronts, DNS moves beyond the idealized conceptual picture, enabling direct observation and quantification of the rate at which water surrounds the larger pore before isolating the remaining air from the external gas phase.

The cumulative effect of the air-entrapment mechanisms described above is reflected in the temporal evolution of the degree of saturation (Figure~\ref{fig_S_t}). During the initial stage of spontaneous imbibition, the degree of saturation increases rapidly, reaching approximately 0.55 after only 12~$\mu$s, compared with an apparent limiting value of about 0.7 by the end of the simulation. At this time, the wetting front has already reached the upper part of the domain locally, particularly along the left side of the pore network, showing that rapid local front propagation does not correspond to uniform volumetric filling of the pore space. However, the saturation rate decreases continuously from the beginning of the process, as the most favourable connected pathways are invaded rapidly. As imbibition progresses, an increasing fraction of the flow occurs through the more viscous water phase, while progressive air entrapment reduces the connected pathways available for further liquid invasion. Together, these effects increase the hydraulic resistance of the remaining flow network and progressively slow further saturation increase. After approximately 30~$\mu$s, the saturation curve is already close to its final value, and only a minor additional increase is observed, although local redistribution of water and slow filling of some larger pore regions may still be ongoing. The resulting flattening of the saturation curve indicates that the majority of the rapid capillary-driven invasion has already taken place, leaving a substantial volume of disconnected air that cannot be readily displaced. The degree of saturation reached at the end of the simulation is therefore not governed solely by the height reached by the wetting front, but by the combined effects of capillary-driven invasion, pore-network connectivity, residual air entrapment, and the persistence of air connected to the atmosphere. The implications of the relatively low attainable degree of saturation and the gradual flattening of the saturation curve are discussed in Section~\ref{sec_discussion}, together with the physical mechanisms and modelling assumptions underlying the present simulations.

\subsubsection{Pressure distribution of trapped air}
\label{subsubsec_air_pressure}

\begin{figure*}
\centering{\includegraphics[width=\textwidth]{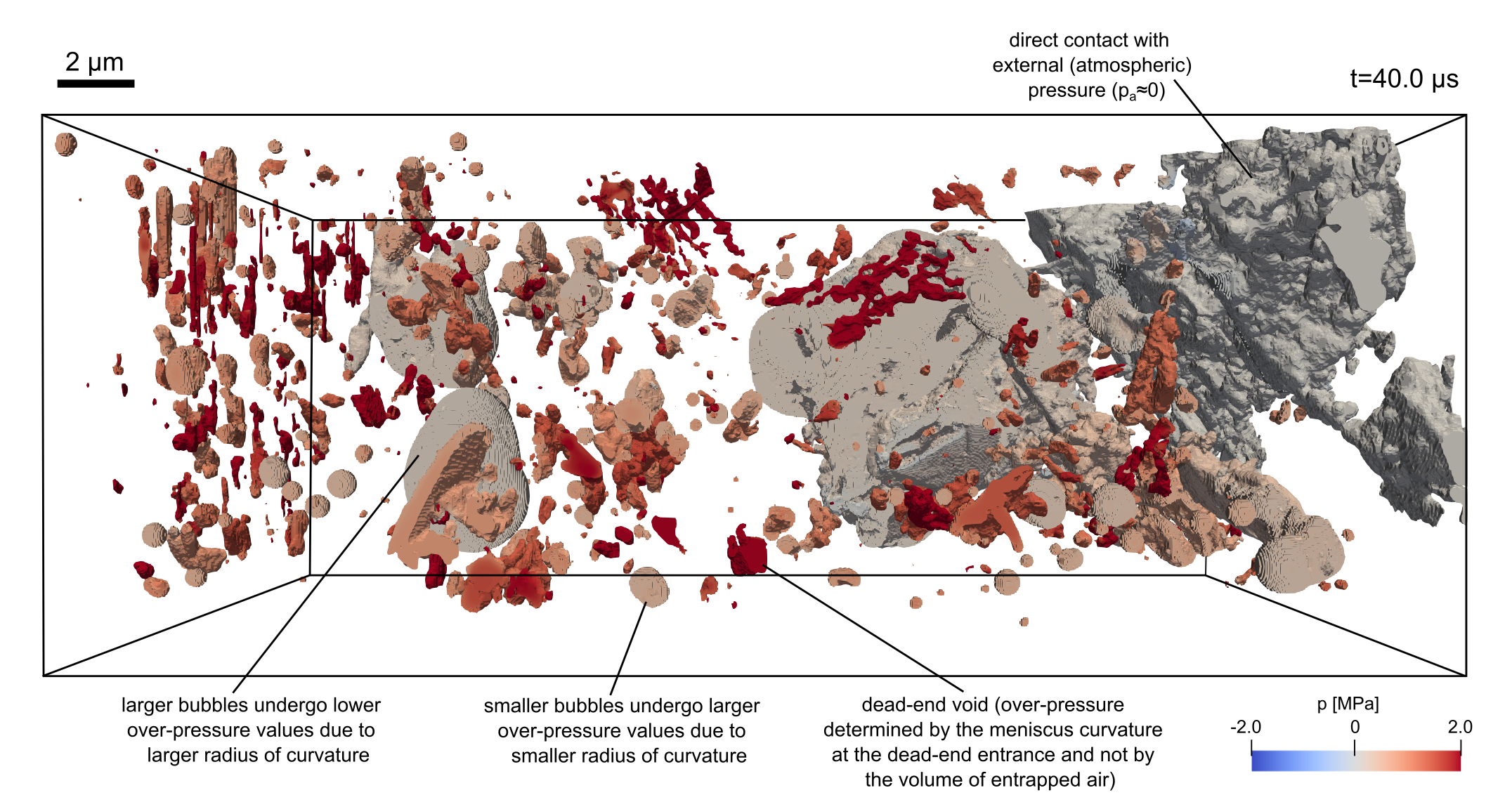}}
\caption{Visualisation of residual air and pressure distribution at \(t=40~\mu\mathrm{s}\). A significant amount of air remains trapped within the pore space despite the prescribed perfectly water-wetting conditions (\(\theta=0^\circ\)). The colours show the pressure distribution within the air phase. Outlet-connected air remains close to the atmospheric reference pressure (\(p_a\approx0\)), whereas isolated trapped-air volumes develop positive overpressures determined by the surrounding water pressure and the curvature of the confining air--water interface. The complete animation is provided in the Supplementary Materials (anim-03).}\label{fig_airPressure}
\end{figure*}

Following air entrapment, each isolated trapped-air volume, hereafter referred to as an air cluster, is no longer connected to the external atmosphere and is therefore no longer constrained by the atmospheric boundary condition. By definition, the capillary pressure is the pressure difference across the air--water interface (Eq.~\ref{eq:pc}). Accordingly, the trapped-air pressure is determined by the surrounding water pressure and the capillary pressure across the air--water interface, \(p_a=p_w+p_c\). The Young--Laplace equation relates the pressure difference, \(p_c\), to the interfacial tension and the local curvature of the air--water interface (Eq.~\ref{eq:yle}). The present DNS resolves the pressure field directly, including the capillary pressure jump across the air--water interface, since the physics described by the Young--Laplace equation is inherently incorporated through the surface-tension force term (Eq. \ref{eq:CSF}) in the governing equations. Figure~\ref{fig_airPressure} presents the resulting pressure distribution within the residual air phase at \(t=40~\mu\mathrm{s}\).

\begin{figure*}
\centering
\includegraphics[width=0.9\textwidth]{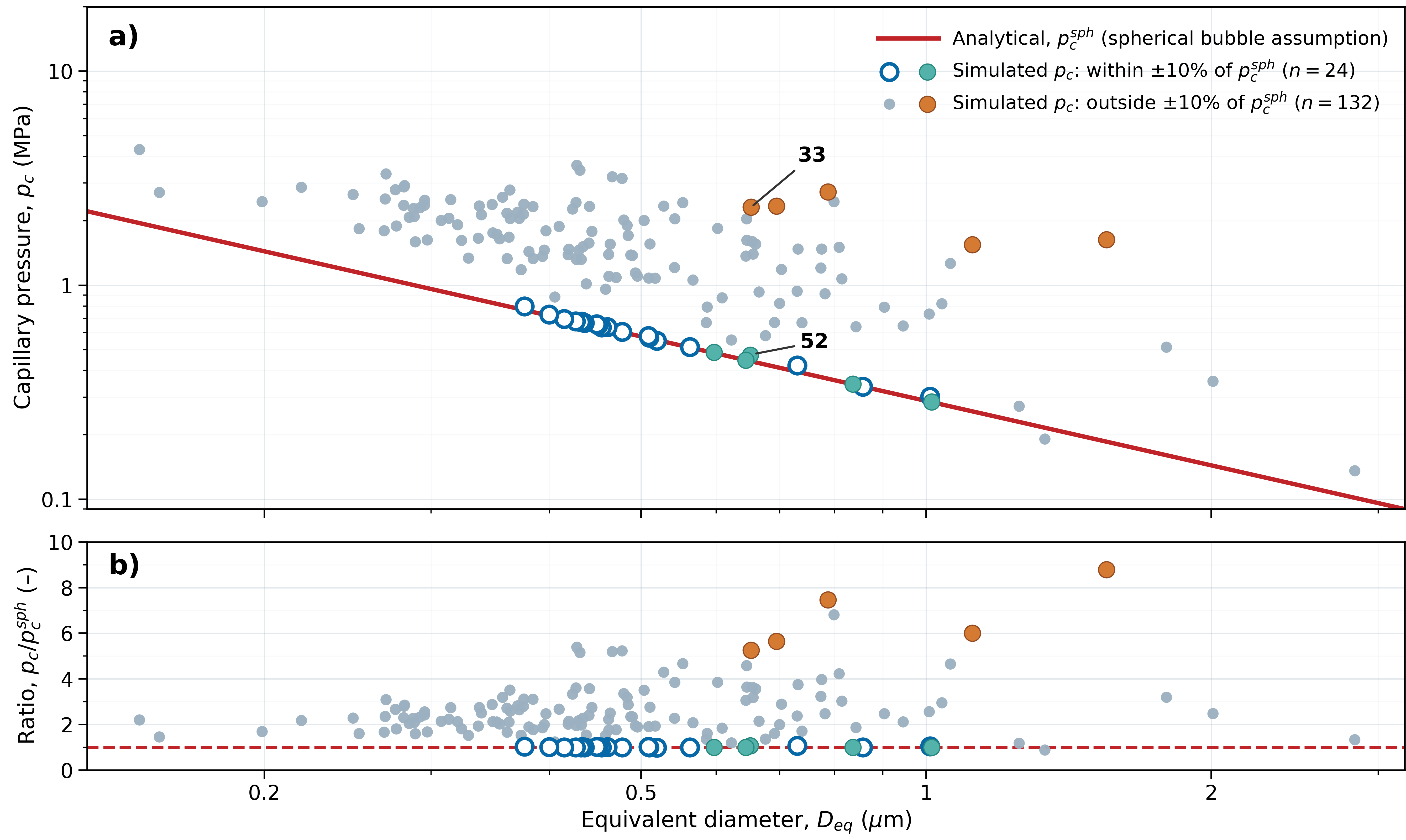}
\caption{Capillary pressure of 156 isolated air clusters at \(t=40~\mu\mathrm{s}\), shown as a function of the volume-equivalent diameter, \(D_{\mathrm{eq}}\) (diameter of a sphere of equal volume): a) DNS-computed \(p_c\) (circles) compared with the analytical capillary pressure assuming a spherical bubble, \(p_c^{\mathrm{sph}}=4\sigma/D_{\mathrm{eq}}\) (red line), and b) their ratio, \(p_c/p_c^{\mathrm{sph}}\). Open blue circles indicate clusters within $\pm10\%$ of $p_c^{\mathrm{sph}}$ ($n=24$)  and filled grey circles outside this range ($n=132$). Five clusters from each group are highlighted in turquoise and orange, respectively, and further examined in Figure~\ref{fig_bubblePressureRend}.}
\label{fig_bubbleGraph}
\end{figure*}

The pressure distribution shown in Figure~\ref{fig_airPressure} is characterised by numerous isolated air clusters exhibiting a wide range of overpressures, with some reaching values on the order of megapascals, while the outlet-connected air phase remains close to the atmospheric reference pressure. This pronounced pressure heterogeneity highlights how local pore-scale conditions can produce markedly different air pressure values. To quantify these differences and examine how they relate to the size and geometry of the trapped-air volumes, a detailed analysis of the isolated air phase was performed. A connected-component analysis was performed to identify distinct air regions at \(t=40~\mu\mathrm{s}\). Regions connected to the outlet, and therefore remaining close to the atmospheric reference pressure, were excluded, along with regions intersecting the other domain boundaries and the smallest air regions. In addition, clusters exhibiting pressure variations greater than 1\% within the trapped-air core region were omitted to restrict the analysis to approximately stable clusters with nearly uniform internal pressure and limited ongoing hydrodynamic or interfacial deformation. These filtering criteria also reduce possible numerical uncertainty in the determination of cluster pressure and volume. The resulting set of 156 well-resolved, isolated, and approximately stable air clusters was retained for the quantitative analysis presented in Figure~\ref{fig_bubbleGraph}.

To examine the effects of trapped-air volume and interface geometry independently of variations in the surrounding water pressure, we shift our focus from the trapped-air pressure \(p_a\) shown in Figure~\ref{fig_airPressure} to the capillary pressure \(p_c\). Figure~\ref{fig_bubbleGraph} compares the DNS-computed capillary pressure of the 156 selected air clusters with the analytical prediction for spherical bubbles. For each cluster, the volume-equivalent diameter, \(D_{\mathrm{eq}}\), is defined as the diameter of a sphere having the same volume as the corresponding trapped-air volume. Using the Young--Laplace relation (Eq.~\ref{eq:yle}), the corresponding analytical capillary pressure for a spherical bubble is \(p_c^{\mathrm{sph}}=4\sigma/D_{\mathrm{eq}}\). As shown in Figure~\ref{fig_bubbleGraph}a, the DNS-computed capillary pressures exhibit substantial scatter relative to the analytical prediction, with clusters of similar equivalent diameter displaying markedly different values. While some clusters closely follow the spherical approximation, many exhibit considerably higher capillary pressures than predicted for a sphere of equal volume. This behaviour is further highlighted in Figure~\ref{fig_bubbleGraph}b, where the ratio \(p_c/p_c^{\mathrm{sph}}\) reveals substantial deviations across a broad range of equivalent diameters. Only 24 of the 156 clusters exhibit capillary pressures within \(\pm10\%\) of \(p_c^{\mathrm{sph}}\), whereas the remaining 132 lie outside this range. These results demonstrate that trapped-air volume alone does not provide a reliable basis for predicting capillary pressure in the investigated pore network, with the spherical approximation frequently underestimating the DNS-computed values.

\begin{figure*}
\centering{\includegraphics[width=\textwidth]{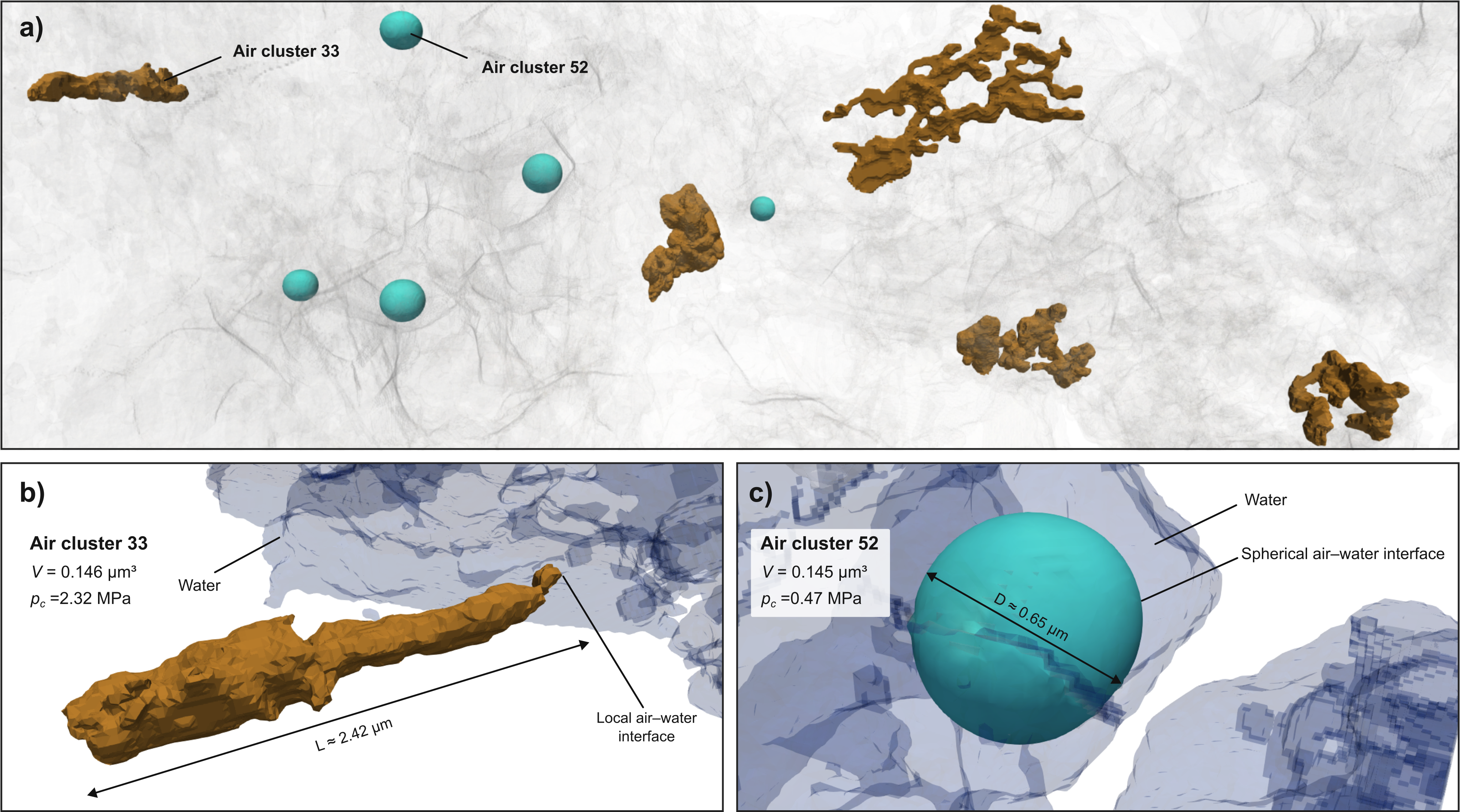}}
\caption{Comparison of selected trapped-air clusters from the two groups identified in Figure~\ref{fig_bubbleGraph}. a) Three-dimensional rendering of five clusters within \(\pm10\%\) of \(p_c^{\mathrm{sph}}\) (turquoise) and five clusters outside this range (orange). b) Cluster 33, an elongated cluster trapped in a dead-end pore, whose overall shape is constrained by the surrounding solid geometry (\(V=0.146~\mu\mathrm{m}^3\) and \(p_c=2.32~\mathrm{MPa}\)); the capillary pressure is governed by the strongly curved local air--water interface rather than by the overall cluster volume. c) Cluster 52, a nearly spherical air bubble predominantly surrounded by water, with a similar volume (\(V=0.145~\mu\mathrm{m}^3\)) but much lower capillary pressure (\(p_c=0.47~\mathrm{MPa}\)), consistent with its approximately spherical air--water interface.}\label{fig_bubblePressureRend}
\end{figure*}

To further investigate the geometrical origin of the deviations observed in Figure~\ref{fig_bubbleGraph}, five clusters from each group were selected and visualised in Figure~\ref{fig_bubblePressureRend}a. The turquoise clusters, whose capillary pressures remain within \(\pm10\%\) of \(p_c^{\mathrm{sph}}\), are indeed nearly spherical, whereas the orange clusters exhibit more irregular and geometrically complex configurations. These contrasting morphologies suggest that the deviations from the spherical approximation are associated with geometrical characteristics that cannot be captured by the volume-equivalent diameter alone.

Clusters 33 and 52 provide a particularly clear comparison (Figures~\ref{fig_bubblePressureRend}b and ~\ref{fig_bubblePressureRend}c). Despite having nearly identical volumes, \(V=0.146~\mu\mathrm{m}^3\) and \(0.145~\mu\mathrm{m}^3\), respectively, their capillary pressures differ by almost a factor of five, with \(p_c=2.32~\mathrm{MPa}\) for cluster 33 and \(p_c=0.47~\mathrm{MPa}\) for cluster 52. Cluster 52 forms a nearly spherical bubble, predominantly surrounded by water, so that its interface curvature is closely related to its volume-equivalent diameter, consistent with the analytical spherical-bubble approximation. In contrast, cluster 33 is elongated within a dead-end pore, with much of its trapped-air volume confined by the solid boundaries and only a limited portion of its boundary forming an air--water interface. The meniscus spanning the pore entrance is strongly curved, resulting in a substantially higher capillary pressure than that of cluster 52, despite their nearly identical volumes. This demonstrates that the capillary pressure can be governed by the local curvature of a relatively small confining meniscus, rather than by the overall dimensions or volume of the enclosed air.

More generally, the results demonstrate that neither the characteristic pore size nor the volume of trapped air uniquely determines the capillary pressure. Instead, the relevant quantity is the curvature of the confining air--water interface, which depends on its specific configuration within the surrounding pore geometry. These observations provide a physical explanation for the substantial variations in capillary pressure revealed by Figure~\ref{fig_bubbleGraph} and contribute to understanding the highly heterogeneous trapped-air pressure distribution shown in Figure~\ref{fig_airPressure}.

The results presented above reveal substantial capillary pressures within trapped-air clusters, reaching several megapascals owing to the strongly curved air--water interfaces at micrometre and sub-micrometre scales. However, the actual pressure within trapped air also depends on the surrounding water pressure (\(p_a=p_w+p_c\), Eq.~\ref{eq:pc}). During active imbibition, the advancing menisci can generate substantial negative gauge pressures within the water phase, potentially offsetting part of the capillary pressure and resulting in considerably lower trapped-air pressures. Towards the end of the simulated imbibition process, when the saturation rate becomes very small, the surrounding water pressure may approach the atmospheric reference pressure, so that the trapped-air overpressure also approaches the capillary pressure. This distinction is important because the actual gas pressure, rather than the capillary pressure alone, plays a central role in the subsequent compression and dissolution of trapped air.

It should be noted, however, that in the present simulations the gas phase is treated as incompressible. The high pressures obtained for some trapped-air clusters therefore reflect the simulated interface configurations, without accounting for changes in gas volume due to compression. When air becomes disconnected from the external atmosphere, its initial pressure reflects the local gas-pressure conditions immediately before entrapment, which are generally close to atmospheric under the conditions considered here. For a compressible gas, the subsequent pressure and volume evolution would depend on the amount of gas initially trapped, its initial pressure, the evolving surrounding water pressure, and the geometry of the confining interface. Gas compression could therefore reduce the trapped-air volume and induce further interface reconfiguration, modifying the local curvature and associated capillary pressure. Therefore, the present results provide insight into the formation of trapped air and its associated pressure fields, but do not capture the coupled pressure--volume evolution associated with gas compressibility or the reduction of trapped-air volume through dissolution. The implications of these assumptions for the subsequent evolution of trapped air are discussed further in Section~\ref{sec_discussion}.

\subsection{Mechanistic analysis using a simplified 2D pore geometry}
\label{subsec_2D}

\subsubsection{Interface evolution and pressure development in an irregular pore geometry}
\label{subsec_2D_results}

While three-dimensional simulations provide a detailed and physically realistic representation of the imbibition process, the simultaneous presence and interaction of numerous evolving fluid--fluid interfaces and pore-scale mechanisms make it difficult to distinguish the contribution of individual mechanisms to the overall imbibition behaviour. Therefore, a two-dimensional pore geometry extracted from a representative slice of the reconstructed FIB--SEM microstructure is analysed in this section. Although such representation simplifies the geometric complexity of the pore space, it enables the fundamental pore-scale mechanisms to be more clearly visualized, isolated, and interpreted before examining how they are modified within realistic three-dimensional pore networks.

\begin{figure*}
\centering{\includegraphics[width=\textwidth]{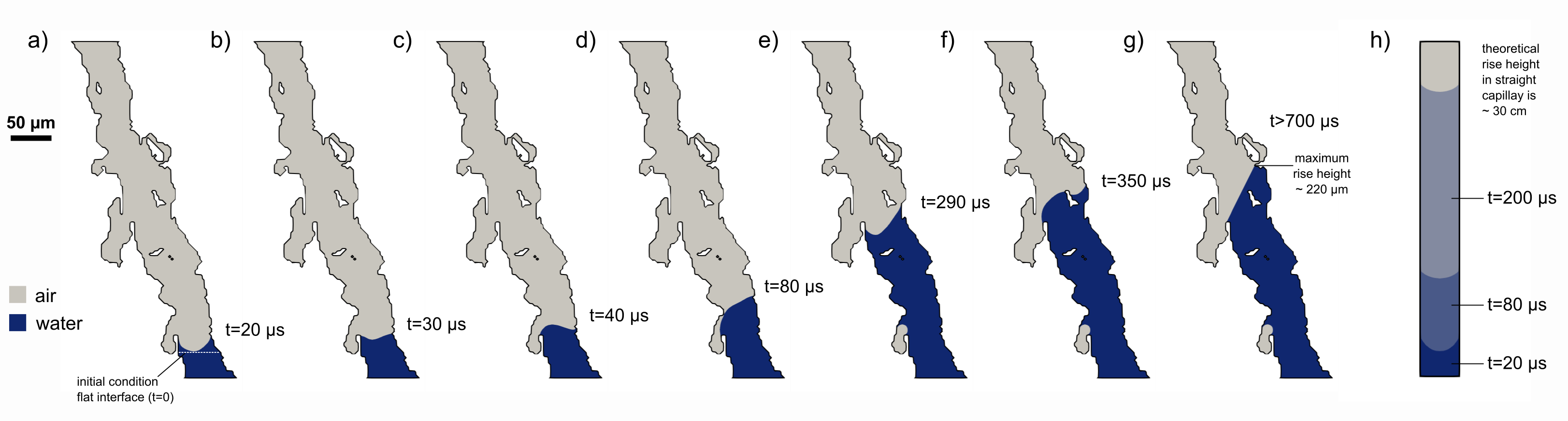}}
\caption{Transient evolution of spontaneous imbibition through the complex 2D capillary. The capillary shape is extracted from a FIB-SEM geometry (Fig.~\ref{fig_geometry}) and uniformly enlarged by a factor of 100 to facilitate identification of the governing pore-scale mechanisms. As the meniscus advances through the irregular pore geometry, it continuously changes shape and curvature, producing strongly non-uniform interface motion. For comparison, in an idealized capillary of similar width (\(\sim 50\,\mu\mathrm{m}\), case h), the advancing meniscus remains concave throughout imbibition, and the theoretical capillary rise (\(\sim 30\,\mathrm{cm}\)), predicted by Jurin's law (Eq.~\ref{eq:jurin}), is approximately three orders of magnitude greater than the height of the simulated domain (\(\sim 0.4\,\mathrm{mm}\)), allowing water to readily reach the top. In the irregular geometry, however, interface advancement is arrested at a geometrically induced capillary barrier, with the final configuration (g) representing the maximum rise height. This demonstrates that pore geometry alone can arrest spontaneous imbibition far below the theoretical rise predicted for an idealized capillary. The complete animation is provided in the Supplementary Materials (anim-04).}\label{fig_2D_part_transient}
\end{figure*}

Figure \ref{fig_2D_part_transient}a-g presents the temporal evolution of spontaneous imbibition through the extracted two-dimensional capillary. The geometry used was uniformly enlarged by a factor of 100 to facilitate a clearer identification and interpretation of the underlying pore-scale mechanisms governing spontaneous imbibition. The role of geometric scale is considered separately in the following section, where the rationale for the present scaling becomes apparent. 

For comparison, Figure~\ref{fig_2D_part_transient}h shows spontaneous imbibition in an idealized two-dimensional capillary of constant width, comparable to the characteristic width of the extracted irregular capillary, and provides a reference for assessing the influence of pore-geometry complexity. Although the meniscus curvature can vary owing to dynamic effects \citep{malenica2024towards}, the advancing meniscus remains concave throughout imbibition in this idealized geometry. The extracted pore geometry, however, forces the interface to undergo rapid reconfigurations as it encounters successive pore constrictions and expansions. Consequently, both the interface shape and its local curvature evolve continuously during spontaneous imbibition, producing complex capillary-pressure distributions that cannot be represented by a single characteristic pore size.

\begin{figure*}
\centering{\includegraphics[width=\textwidth]{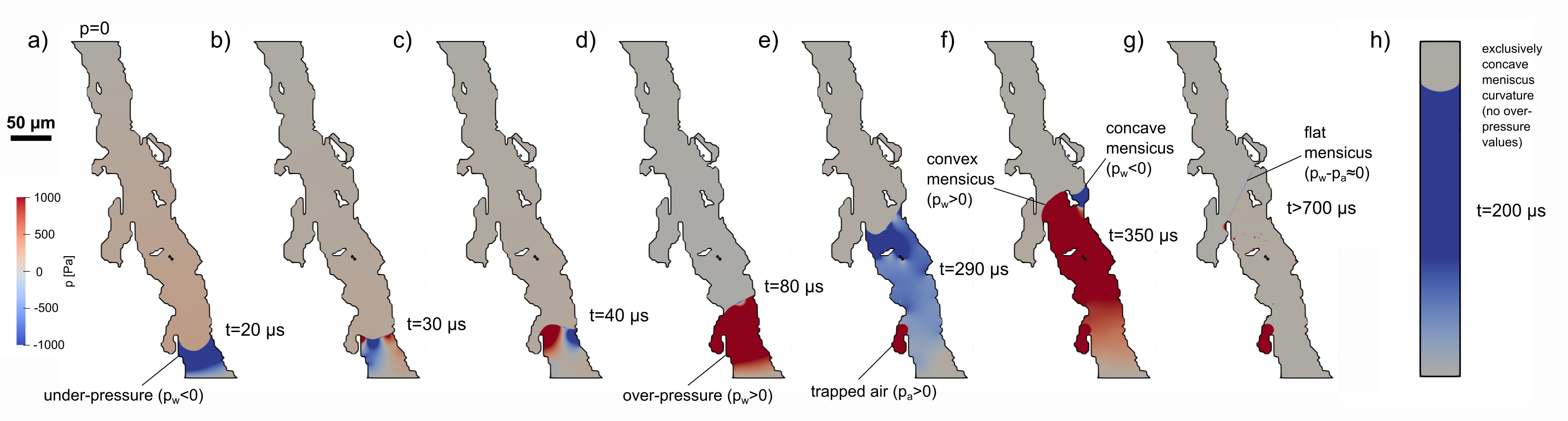}}
\caption{Pressure distribution corresponding to the air-water distributions shown in Fig.~\ref{fig_2D_part_transient}. In the idealized geometry (case h), the water phase remains under negative pressure. In the more complex geometry, meniscus reconfigurations generate large pressure variations, including both negative and positive water pressures. Reconfigurations with positive water pressure (i.e., negative capillary pressure; $p_c=p_a-p_w \approx -p_w$) slow imbibition, while pore-scale inertia can enable the interface to overcome this opposing capillary force. Case (g) represents the final state, where inertia is no longer sufficient to overcome the capillary barrier and the nearly flat interface corresponds to approximately zero capillary pressure, leaving no further capillary driving force. The air phase above the meniscus remains close to atmospheric pressure, whereas entrapped air develops much larger overpressures determined by the surrounding water pressure and air-water interface curvature. The pressure color scale is symmetric about zero to distinguish under- and over-pressure values. The complete animation is provided in the Supplementary Materials (anim-04).}\label{fig_2D_part_transient_pressure}
\end{figure*}

To obtain a deeper understanding of this process, Figure~\ref{fig_2D_part_transient_pressure} presents the corresponding pressure distributions in both fluid phases at the same time instants shown in Figure~\ref{fig_2D_part_transient}. The pressure colour scale spans equal negative and positive pressure ranges, allowing underpressure and overpressure regions to be clearly distinguished.

We first consider the pressure distribution in the air phase. As the advancing water front displaces air towards the outlet, a pressure gradient develops within the connected air phase. The pressure is highest immediately ahead of the advancing meniscus and decreases towards the outlet, where atmospheric pressure is imposed. The low viscosity of air contributes to the small pressure gradient required to sustain the airflow, particularly at the relatively low velocities considered here. The gradient is largest during the initial inertial stage (Fig.~\ref{fig_2D_part_transient_pressure}a), when the interface velocity is highest. As the pore space progressively fills with water, an increasing fraction of the flow occurs through the more viscous water phase, increasing viscous dissipation and slowing the interface. The required air-pressure gradient therefore decreases, and the connected air phase remains close to atmospheric pressure throughout most of the imbibition process.

A different situation arises when air becomes entrapped. Figure~\ref{fig_2D_part_transient_pressure}e shows an isolated air pocket within a dead-end pore. In this case, the trapped-air pressure is no longer governed by the atmospheric outlet boundary condition. Instead, it is determined by the surrounding water pressure and the capillary pressure across the enclosing air--water interface, as discussed previously for the three-dimensional simulations (Section~\ref{subsubsec_air_pressure}). Owing to the relatively small radius of curvature of the entrapped meniscus, the dead-end pore develops a pronounced trapped-air overpressure, which also responds directly to changes in the surrounding water pressure or the meniscus curvature.

We now turn our attention to the water-pressure field. Unlike the connected air phase, where pressure variations remain small because of the low viscosity of air and its connection to the atmospheric outlet, the water-pressure field varies strongly as the meniscus reconfigures through the irregular pore geometry. Changes in interface curvature modify the capillary pressure according to the Young–Laplace equation, producing pronounced spatial and temporal variations in water pressure. Fig.~\ref{fig_2D_part_transient_pressure} illustrates this behaviour through several characteristic stages of interface evolution.

Initially (Figure~\ref{fig_2D_part_transient_pressure}a), the meniscus is concave, producing a positive capillary pressure and consequently underpressure within the water phase. Spontaneous imbibition proceeds as the positive capillary pressure drives the water phase upward.

As the interface encounters the entrance to the large dead-end pore on the left (Figures~\ref{fig_2D_part_transient_pressure}b and c), the already non-uniform interface advancement becomes substantially more pronounced owing to temporary contact-line pinning at the abrupt pore expansion. Consequently, different parts of the meniscus undergo different local reconfigurations and develop substantially different local curvatures. Accordingly, concave regions maintain positive capillary pressure, whereas locally convex regions progressively develop negative capillary pressure. As discussed above, the connected air pressure may be approximated as atmospheric (i.e. \(p_a \approx 0\) in terms of relative pressure). Consequently, local variations in capillary pressure are reflected almost directly in the water pressure immediately beneath the meniscus (\(p_w \approx -p_c\)), whereas farther from the interface the water-pressure field remains primarily controlled by the pressure imposed at the interface, with additional contributions from flow-induced pressure gradients and gravity. As a result, neighbouring regions of the water phase simultaneously exhibit positive and negative pressure values, giving rise to local pressure gradients and a continuous redistribution of flow within the liquid. Regions of positive capillary pressure continue to promote spontaneous imbibition, whereas regions of locally negative capillary pressure generate pressure forces opposing the overall direction of interface advancement.

As the interface advances further (Figure~\ref{fig_2D_part_transient_pressure}d), an increasing portion of the meniscus becomes locally convex and the opposing capillary-pressure region expands. This is reflected in the water-pressure field, where the convex interface produces water pressures above atmospheric pressure ($p_w>0$). Continued interface advancement is therefore possible only because the inertia of the moving water phase, manifested through its momentum and kinetic energy, is sufficient to overcome this temporarily unfavourable capillary configuration \citep{pavuluri_towards_2020,malenica2024towards}. Once the contact line depins, the meniscus reconfigures again into a favourable concave shape, restoring positive capillary pressure and allowing spontaneous imbibition to resume (Figure~\ref{fig_2D_part_transient_pressure}e).

As spontaneous imbibition proceeds further (Figure~\ref{fig_2D_part_transient_pressure}f), the advancing interface becomes separated into two simultaneously evolving menisci. Each meniscus undergoes its own local reconfiguration, developing distinct local curvatures and therefore different capillary pressures. Consequently, the water-pressure field becomes increasingly heterogeneous, giving rise to progressively more complex local pressure gradients and flow redistribution. In this particular case, most of the water phase is characterised by positive pressure, while only a small region adjacent to one of the menisci remains under negative pressure. This behaviour reflects the distinct contribution of each meniscus to the local pressure field, with the overall pressure distribution resulting from their simultaneous evolution.

Eventually, the two advancing menisci coalesce into a single interface spanning the pore expansions on both sides of the solid wall (Figure~\ref{fig_2D_part_transient_pressure}g). This interface reaches a geometrical configuration in which the capillary barrier can no longer be overcome. Unlike the preceding pore expansions, the combined effect of the increasingly unfavourable interface configuration and the reduced kinetic energy of the moving water phase, resulting from continuous viscous dissipation, prevents the final capillary barrier from being overcome. As the interface becomes nearly flat, its curvature approaches zero and the capillary pressure vanishes, eliminating the driving force for further spontaneous imbibition. Consequently, the contact line remains pinned and the interface reaches its final equilibrium configuration. Further advancement would require the meniscus to become convex, corresponding to negative capillary pressure. Such a configuration is energetically unfavourable under the imposed complete-wetting conditions, making the pinned interface the energetically preferred configuration.

\begin{figure*}
\centering{\includegraphics[width=\textwidth]{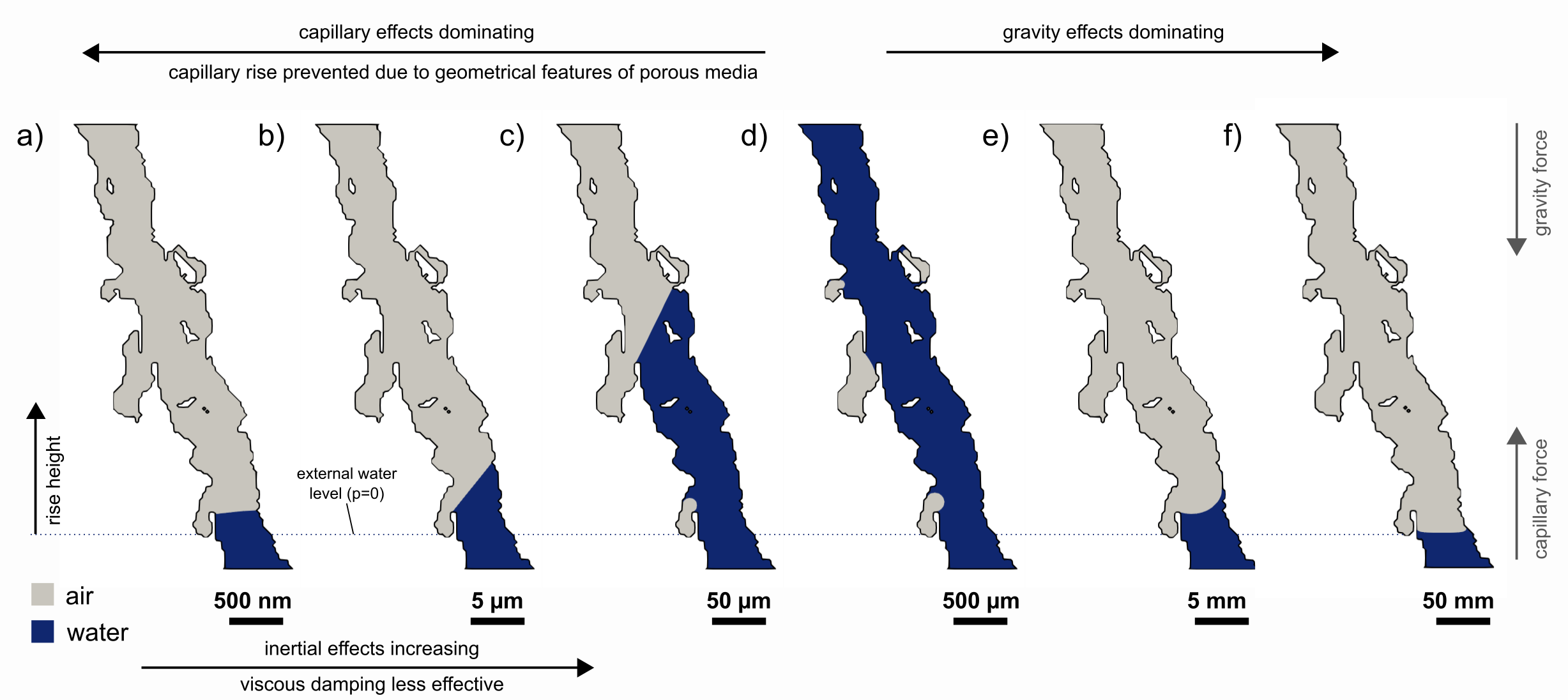}}
\caption{Maximum water rise at different pore-geometry scales. The same pore geometry is uniformly scaled by a factor of 10 from left to right, with case (a) corresponding to the original FIB-SEM obtained size. At the smallest scales, spontaneous imbibition is arrested at geometrically induced capillary barriers because pore-scale inertia is insufficient to overcome abrupt pore expansions, despite the highest theoretical equilibrium capillary rise. At intermediate scales, reduced viscous resistance allows sufficient momentum to overcome successive barriers. At the largest scales, gravity becomes dominant relative to capillary forces and limits further rise. The results demonstrate a scale-dependent transition in the mechanism limiting spontaneous imbibition. The complete animation is provided in the Supplementary Materials (anim-05).}\label{fig_2D_part_final_state}
\end{figure*}

This behaviour differs fundamentally from that of the idealized capillary (Figure~\ref{fig_2D_part_transient_pressure}h), where the meniscus remains concave throughout the imbibition process and the capillary pressure therefore remains positive. The corresponding theoretical equilibrium rise is approximately 30~cm, whereas in the irregular geometry spontaneous imbibition is arrested at a maximum rise height of only about \(220~\mu\mathrm{m}\).

Overall, these results demonstrate that capillary pressure in irregular pore geometries cannot generally be represented by a single characteristic value derived from pore size, but varies continuously along the evolving interface. This behaviour departs from the simplified capillary representation underlying the analytical models discussed in Section~\ref{subsec_physics_SI}.

\subsubsection{Influence of pore size on the dominant imbibition mechanism}
\label{subsec_size_effects}

The previous results demonstrate that spontaneous imbibition in irregular pore geometries is governed by the competition between capillary forces, viscosity, and inertia. The relative importance of these mechanisms, together with gravity, changes with pore size. To investigate this scale dependence, the extracted capillary geometry was uniformly scaled by a factor of 10 in both spatial directions while preserving its shape. Six geometries spanning five orders of magnitude in characteristic size were simulated under otherwise identical fluid properties and boundary conditions. 

For reference, the equilibrium rise in an idealized two-dimensional capillary of uniform width $w$ can be approximated using Jurin's law (Section~\ref{subsec_physics_SI}) as $h \approx 1.5\times10^{-5}/w$, where both $h$ and $w$ are expressed in metres. This corresponds to heights ranging from approximately 30 m to 0.3 mm across the considered characteristic pore widths, neglecting potential cavitation at the smallest scales and gravitational meniscus deformation at the largest scales. These values provide a theoretical reference rather than a direct prediction of the maximum rise in the irregular pore geometry.

Figure~\ref{fig_2D_part_final_state} presents the final air-water distributions obtained for the different spatial scales. At the smallest scale (Figure~\ref{fig_2D_part_final_state}a), corresponding to the original geometry extracted from the FIB-SEM reconstruction, the interface becomes arrested at the first pronounced pore expansion because the inertia of the advancing liquid is insufficient to overcome the geometrically induced capillary barrier. This occurs despite a theoretical equilibrium rise of several tens of metres predicted for an idealized capillary of comparable width. As the geometry is uniformly enlarged (Figures~\ref{fig_2D_part_final_state}b-d), inertial forces increase, allowing the interface to overcome successive capillary barriers and progressively advance through the pore until it eventually reaches the outlet (Figure~\ref{fig_2D_part_final_state}d). This increase arises because viscous resistance decreases relative to capillary forces as the characteristic length scale increases, allowing the advancing interface to attain higher velocities and therefore greater momentum before encountering each geometrical widening.

Further increases in pore size (Figures~\ref{fig_2D_part_final_state}e and~\ref{fig_2D_part_final_state}f) no longer promote capillary rise. Although inertial effects remain increasingly important relative to viscous resistance, enlarging the geometry also reduces the capillary pressure, while simultaneously increasing the volume of the water column and therefore the total gravitational body force acting on the liquid. Consequently, gravity eventually becomes comparable to the capillary driving force, and the interface approaches a capillary--gravity equilibrium below the top of the domain, consistent with the equilibrium balance underlying Jurin's law. The dominant mechanism therefore shifts from local geometrically induced capillary barriers to a global balance between capillary and gravitational forces.

\begin{figure*}[!t]
\centering{\includegraphics[width=\textwidth]{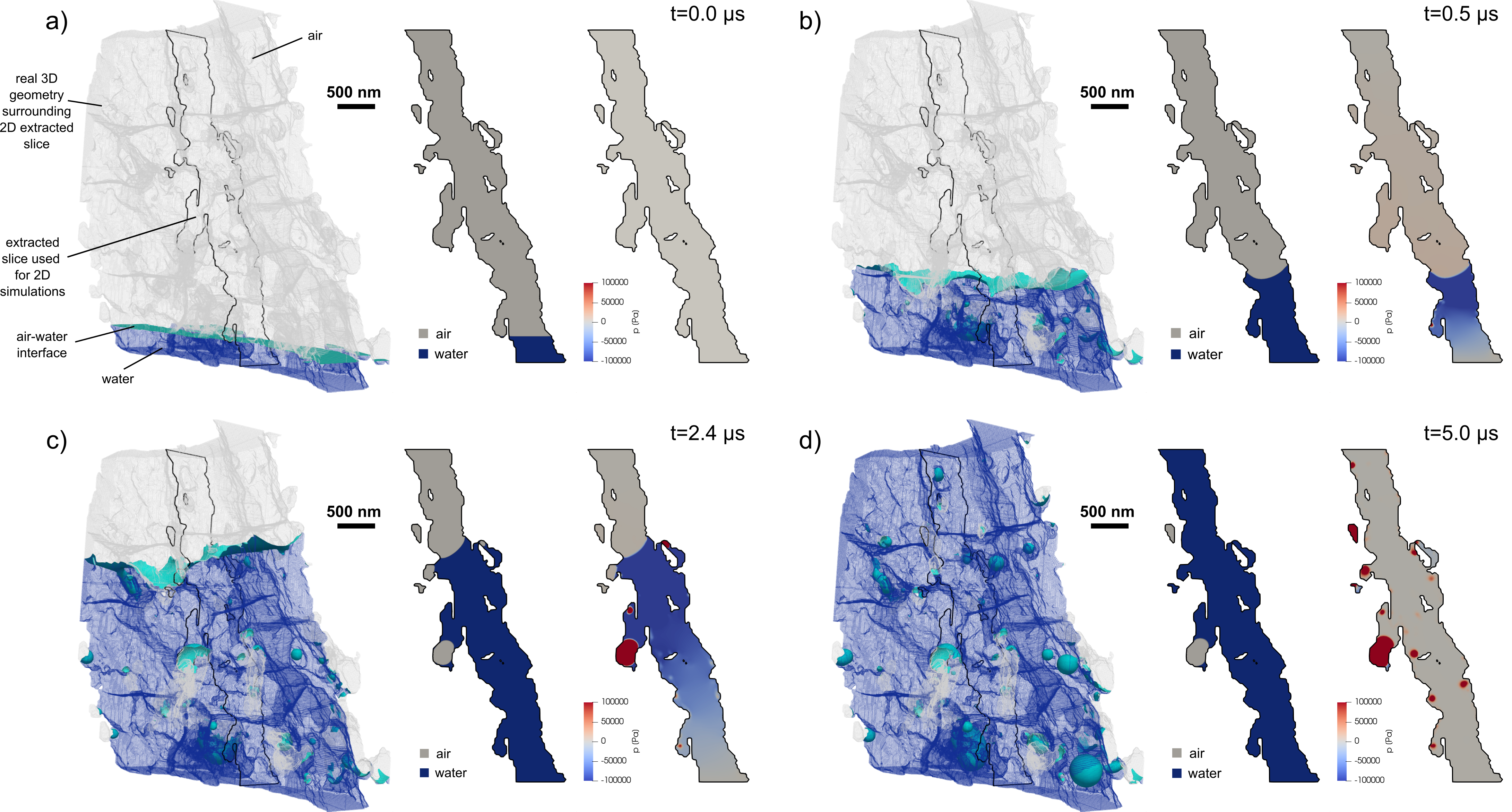}}
\caption{Capillary rise through the 3D micro-crack surrounding the cross-section used in the preceding 2D simulations, here considered at its original FIB-SEM scale (case shown in Figure~\ref{fig_2D_part_final_state}a). The accompanying 2D slices show the air--water distributions and pressure fields extracted from the 3D simulation. Whereas spontaneous imbibition was arrested in 2D (Figure~\ref{fig_2D_part_final_state}a), the additional geometrical complexity of the 3D pore space allows water to readily reach the top of the domain. The three-dimensional interface morphology allows the total curvature, and therefore the capillary pressure, to remain positive even when the meniscus appears locally convex in the extracted 2D slice. The complete animation is provided in the Supplementary Materials (anim-06).}\label{fig_3D_crack}
\end{figure*}

For the isolated pore geometry considered here, the results reveal three distinct spontaneous imbibition regimes governed by pore scale. At the smallest scales, capillary rise is arrested by geometrically induced capillary barriers because pore-scale inertia remains insufficient, partly owing to strong viscous damping of the advancing flow. At intermediate scales, reduced relative viscous resistance allows greater momentum to develop, enabling the interface to overcome successive capillary barriers while gravitational effects remain weak, resulting in the greatest rise relative to the pore-geometry size. At the largest scales, gravity progressively becomes the dominant limiting mechanism, preventing further capillary rise despite the increasing importance of inertia.
These findings demonstrate that the dominant mechanism governing spontaneous imbibition is inherently scale dependent and cannot be inferred solely from the magnitude of capillary pressure.

While the isolated two-dimensional geometry provides valuable insight into the mechanisms governing interface reconfiguration, pressure evolution, capillary barriers, and pore-scale inertia, it represents a simplified system with a strongly constrained flow topology. Once the advancing interface becomes arrested by a capillary barrier, the wetting front cannot redistribute through alternative connected pathways. The following section therefore investigates how these same pore-scale mechanisms manifest in realistic three-dimensional cementitious microstructures.

\subsection{From isolated 2D pore geometry back to realistic 3D pore space}
\label{subsec_back_to_3D}

Figure~\ref{fig_3D_crack} compares spontaneous imbibition in the original three-dimensional pore space and the isolated two-dimensional geometry analysed in the previous section, with the highlighted slice showing the air--water distribution and pressure fields extracted from the three-dimensional simulation along the corresponding cross-section. Although the surrounding three-dimensional pore space resembles a thin micro-crack, for which a two-dimensional approximation would intuitively be expected to capture the essential imbibition behaviour, the predicted dynamics differ fundamentally. In contrast to the isolated two-dimensional simulations, where spontaneous imbibition was arrested shortly after the first geometrical widening (Figure~\ref{fig_2D_part_final_state}a), water in the three-dimensional geometry reaches the top of the computational domain, demonstrating that even seemingly quasi-two-dimensional pore geometries may not be represented accurately by an isolated two-dimensional model.

A primary reason for this behaviour lies in the three-dimensional shape of the advancing meniscus. Unlike isolated two-dimensional simulations, where capillary pressure depends only on the curvature of the interface within the plane of the cut, the capillary pressure in three dimensions depends on the total interface curvature, which can be expressed in terms of the two principal curvatures of the interface, as discussed in Section~\ref{subsec_physics_SI}. Consequently, even when the meniscus appears locally convex within the extracted two-dimensional slice, the total three-dimensional curvature may still remain positive owing to the additional curvature of the interface outside the plane of the slice. This can be seen, for example, at \(t=1.5~\mu\mathrm{s}\) in the Supplementary animation, where the meniscus appears locally convex in the extracted slice while the adjacent water remains under negative pressure. As a result, negative capillary-pressure regions responsible for capillary barriers in the isolated two-dimensional simulations do not develop.

\begin{figure*}
\centering{\includegraphics[width=0.85\textwidth]{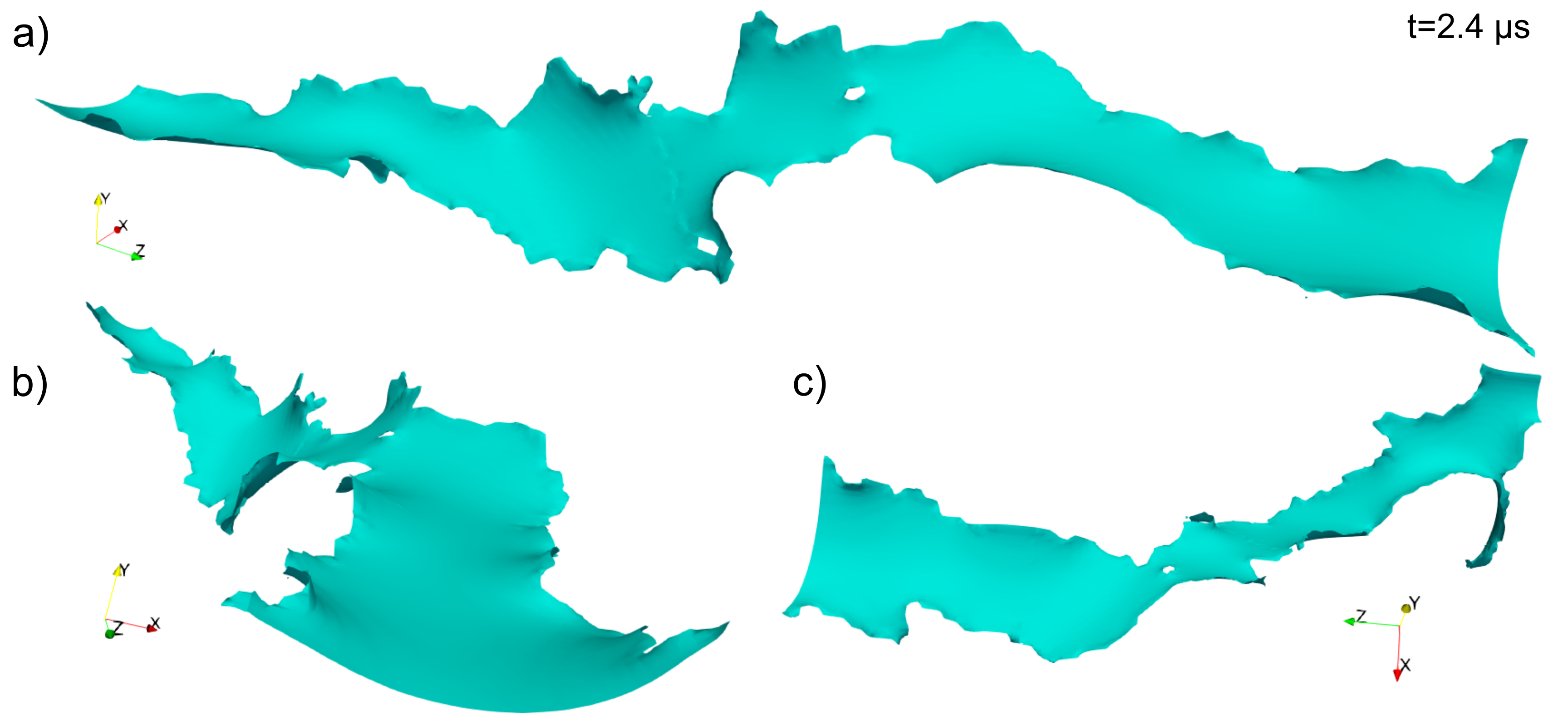}}
\caption{Three-dimensional visualization of the advancing air-water interface shown in Figure~\ref{fig_3D_crack}c. The same meniscus is shown from three different viewing directions to illustrate its complex morphology. Although the surrounding pore space appears quasi-two-dimensional, resembling a thin micro-crack, the meniscus develops pronounced spatial variations in shape and curvature. The figure illustrates the fully three-dimensional nature of the advancing interface and its continuous contact lines, which cannot be captured by an isolated two-dimensional model.}\label{fig_3D_interface}
\end{figure*}

This three-dimensional nature of the advancing interface is further illustrated in Figure~\ref{fig_3D_interface}, which shows the reconstructed air-water interface at a representative instant during spontaneous imbibition. Although the thin, crack-like pore geometry might suggest that a two-dimensional approximation would be sufficient, the advancing meniscus develops a highly irregular geometry, exhibiting pronounced variations in both interface shape and curvature across the width of the crack. If the interface remained unchanged in the third dimension, the two-dimensional approximation would accurately represent the three-dimensional geometry. Instead, the reconstructed meniscus demonstrates that the interface evolves continuously in all three spatial directions, giving rise to the complex distribution of interface curvature responsible for the capillary-pressure variations discussed above. 

Furthermore, in the isolated two-dimensional model, each advancing meniscus is attached to the pore walls through only two contact points. In contrast, the three-dimensional meniscus remains in contact with the surrounding solid along one or more continuous contact lines. Consequently, local contact-line pinning constrains only a limited segment of the advancing interface, while the remaining contact line continues to evolve and deform the meniscus. Because the three-dimensional meniscus forms a continuous interface, its different regions remain capillarily coupled through surface tension: local advancement and deformation of one part of the interface can induce reconfiguration of neighbouring regions, facilitating depinning and sustaining wetting-front advancement.

In addition to the fundamentally different topology of the advancing interface, the three-dimensional pore network exhibits a substantially higher degree of connectivity than its two-dimensional counterpart. Although branching, interface coalescence, and multiple pathways for wetting-front advancement may also occur in two-dimensional geometries, these phenomena become considerably more pronounced in three dimensions owing to the much larger number of possible pore interconnections. Consequently, locally unfavourable interface configurations within individual pores no longer determine the overall advancement of spontaneous imbibition, as neighbouring parts of the pore network may continue to fill simultaneously, allowing the wetting front to redistribute through adjacent pathways. Local geometrical barriers and temporary contact-line pinning therefore exert a much weaker influence on the global imbibition behaviour. Rather than being controlled by the evolution of a comparatively small number of isolated interfaces, spontaneous imbibition is governed by the collective evolution of numerous interconnected interfaces developing throughout the pore network.

\begin{figure*}
\centering{\includegraphics[width=\textwidth]{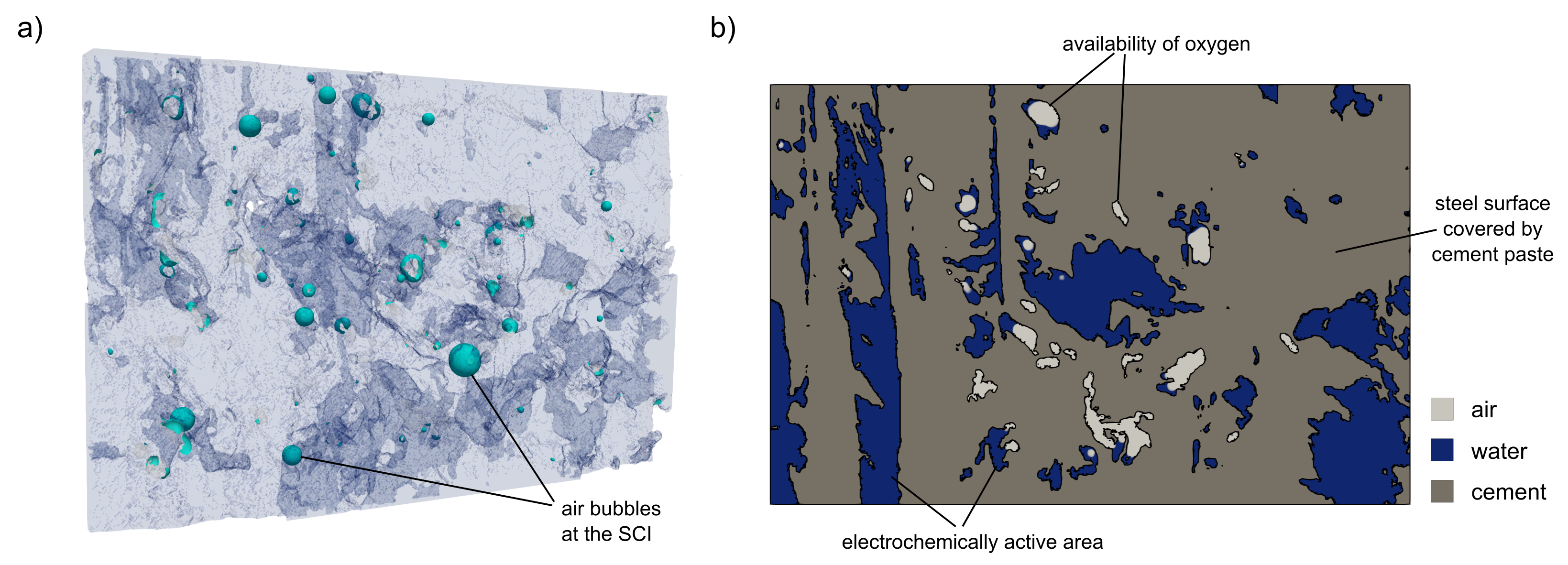}}
\caption{Moisture distribution at the steel-concrete interface (SCI): a) three-dimensional visualization of the air-water distribution and entrapped air surrounding the reinforcing steel; b) distribution of air, water, and cement in direct contact with the steel surface. Water-covered regions indicate the potentially electrochemically active steel area, while adjacent air-filled regions may provide local access to gaseous oxygen, illustrating how pore-scale DNS can directly resolve corrosion-relevant moisture conditions at the steel surface.}
\label{fig_sci}
\end{figure*}

The simplified two-dimensional simulations therefore remain highly valuable because they provide a computationally efficient framework for visualizing, isolating, and understanding the local pore-scale mechanisms underlying spontaneous imbibition. However, their ability to predict the overall imbibition behaviour depends on how well the two-dimensional approximation captures the relevant interface geometry and pore-network connectivity. Realistic three-dimensional pore geometries introduce out-of-plane interface curvature, continuous contact-line dynamics, and highly interconnected flow pathways that can fundamentally modify the resulting imbibition behaviour.


\subsection{Moisture at the steel-concrete interface}
\label{subsec_sci}

While the previous sections focused on understanding the fundamental mechanisms governing spontaneous imbibition, the same pore-scale DNS simulations can also provide direct insight into local moisture conditions at the steel-concrete interface (SCI). Such information is of particular importance because corrosion behaviour cannot be fully characterized solely by the overall degree of saturation. Rather, it depends on the local distribution of water and air surrounding the reinforcing steel, which provides information about potentially electrochemically active steel area and local access to oxygen \citep{stefanoni2018electrochemistry,stefanoni2019kinetics}. Unlike conventional experimental techniques or continuum-scale models, which typically provide only averaged moisture information, interface-resolved DNS simultaneously resolves the evolving air-water distribution, interface geometry, pressure field, and local wetting conditions within realistic cementitious microstructures.

Figure~\ref{fig_sci} presents a detailed view of the steel-concrete interfacial zone extracted from the three-dimensional simulation presented in Section~\ref{subsec_3D_SI}. The analysis focuses on the final air-water distribution surrounding the reinforcing steel and the resulting local wetting conditions at the steel surface. Most of the steel surface remains in direct contact with the solid cementitious matrix rather than the pore space (Figure~\ref{fig_sci}b), reflecting the relatively low resolved porosity of the analysed mortar microstructure (approximately 17\%).

The three-dimensional air-water distribution shown in Figure~\ref{fig_sci}a reveals numerous entrapped air bubbles persisting in the immediate vicinity of the reinforcing steel, producing highly heterogeneous local moisture conditions. Accordingly, Figure~\ref{fig_sci}b demonstrates that, despite the completely hydrophilic conditions assumed in the simulations, only part of the steel surface is wetted by liquid water, while the remaining other pore-exposed regions remain in contact with the air phase. Consequently, only a fraction of the steel surface is potentially electrochemically active, while neighbouring air-filled regions may provide local access to oxygen. This spatial coexistence of water-covered and air-accessible regions creates conditions that are particularly relevant to corrosion processes, illustrating the importance of resolving moisture heterogeneity directly at the steel--concrete interface. In the present simulations, the steel surface is assigned the same equilibrium contact angle (\(\theta=0^\circ\)) as the surrounding cementitious matrix. A distinct steel contact angle was not introduced because the present analysis is intended primarily to demonstrate the capability of pore-scale DNS to resolve SCI moisture conditions rather than to provide a detailed investigation of steel--concrete wettability effects; these will be addressed in future work.

%% file: 04_Implications.tex
\section{Discussion and implications}
\label{sec_discussion}

\subsection{Pore-scale DNS as a virtual experiment}

One of the principal advantages of image-based pore-scale DNS is that it provides direct access to transient processes and physical quantities that remain extremely difficult to resolve experimentally at comparable spatial and temporal scales.  
In this sense, the present framework can be regarded as a virtual pore-scale experiment, in which transient multiphase processes can be interrogated under controlled conditions. The evolving three-dimensional air--water interface can be visualized throughout the entire pore space while simultaneously resolving local phase pressures, interface curvature, contact-line evolution, phase connectivity, and the formation and relaxation of trapped-air volumes. Moreover, the computational geometry can be selectively isolated, simplified, rescaled, or reoriented, while boundary and initial conditions can be systematically varied, allowing the influence of individual physical and geometrical factors to be examined independently while retaining a direct connection to the experimentally reconstructed microstructure. Such controlled variations would be extremely difficult, if not impossible, to realize experimentally. This also provides a direct means of testing the range of validity of simplified pore-scale descriptions and identifying the conditions under which their underlying assumptions begin to break down. This capability was exploited here through complementary analyses of the realistic three-dimensional pore network, isolated two-dimensional geometries, and systematically rescaled pore structures.

The value of this approach lies particularly in the ability to resolve the evolving interface geometry and associated flow fields simultaneously. Interface shape can be related directly to curvature, curvature to capillary pressure, and the resulting pressure field to the local balance of forces governing interface motion and liquid redistribution. These interfacial and flow quantities can therefore be followed simultaneously as the pore-scale configuration evolves, rather than inferred from integral or time-averaged measurements. Similarly, the transient formation of residual air can be followed from the initial invasion of the pore network through progressive gas-phase disconnection to the final trapped-air configuration. Such information remains largely inaccessible at the nanometre-to-micrometre spatial scale and microsecond temporal resolution considered here.

\subsection{Physical interpretation and model scope}

The combined results indicate that spontaneous imbibition in cementitious materials should not be interpreted primarily as the independent filling of pores according to a characteristic pore radius. In idealized capillaries, the pore geometry uniquely constrains the meniscus shape, allowing capillary pressure to be related directly to the capillary dimensions. In irregular pore geometries, this one-to-one relationship breaks down because local constrictions, expansions, branching pathways, and surface irregularities continuously reconfigure the air--water interface. Consequently, capillary pressure varies in space and time throughout the pore network, rather than being represented by a single characteristic value determined by pore size alone.

The resulting interface motion reflects the local competition between capillary, viscous, inertial, and gravitational forces. The geometric-scaling simulations demonstrate that their relative importance changes strongly with pore size. At very small scales, high capillary pressure does not necessarily imply easier propagation because strong viscous damping limits the momentum available for interface reconfiguration and may allow geometrically induced capillary barriers to arrest the advancing meniscus.
At intermediate scales, reduced relative viscous resistance permits greater liquid momentum and facilitates passage through such barriers. At sufficiently large scales, the reduction in capillary pressure together with the increasing gravitational force causes the interface to approach a capillary--gravity equilibrium. Thus, neither capillary pressure nor pore size alone determines the imbibition dynamics; the relevant behaviour emerges from the scale-dependent balance between the governing forces.

This force balance must furthermore be interpreted within the full three-dimensional pore topology. Local capillary barriers and contact-line pinning observed in two dimensions remain physically relevant, but their influence on the overall wetting process is substantially reduced when neighbouring interfaces can continue to advance and redistribute through alternative connected pathways. Branching of the wetting front, coalescence and reconnection of neighbouring interfaces, and simultaneous invasion through multiple pathways further contribute to this collective network behaviour. Three-dimensional interface curvature further modifies the local capillary-pressure field because the out-of-plane curvature may differ substantially from the in-plane curvature and can therefore alter the magnitude and even the sign of the total interface curvature. The mechanisms identified in simplified two-dimensional geometries therefore remain useful for understanding individual pore-scale events, but their manifestation in realistic materials is controlled by multidimensional interface reconfiguration and pore-network connectivity. These limitations should, therefore, be kept in mind when interpreting results from two-dimensional simulations, particularly when extrapolating their findings to realistic three-dimensional pore networks.

Air entrapment forms an integral part of this same dynamic network process and can arise in several distinct but related forms. These include bypass and subsequent reconnection of wetting fronts around slowly filling pore bodies, sealing of dead-end pores by advancing interfaces, and the formation of both detached air bubbles and air volumes that remain attached to the solid skeleton, where geometrical constraints and contact-line pinning prevent detachment from the solid surface even under perfectly hydrophilic conditions. The conceptual mechanism proposed by Fagerlund \citep{fagerlund2004service} therefore represents one characteristic trapping pathway within a broader network process in which local capillary suction, interface motion, pore connectivity, and gas-phase disconnection evolve simultaneously.

Once the gas phase becomes isolated, its behaviour also changes qualitatively. While connected to the external atmosphere, the low viscosity of air implies that only small pressure gradients are required to sustain its displacement, so that the gas pressure remains close to atmospheric pressure throughout most of the connected air phase. Once trapped, however, gas volumes are no longer constrained by the atmospheric boundary condition. Their pressure is then determined by the surrounding water pressure together with the capillary-pressure jump across the enclosing interface, and strongly curved micrometre- and sub-micrometre-scale menisci may generate substantial overpressures.

This trapping behaviour also helps explain the continuous reduction in the saturation rate observed in the three-dimensional simulations. Rapid advancement through the most favourable connected pathways occurs early, while substantial parts of the pore space remain only partially invaded or air-filled. As imbibition progresses, further water redistribution becomes increasingly controlled by slower-filling regions and the progressive isolation of gas within the pore network, which increasingly restricts further liquid invasion and contributes to the reduction in the imbibition rate. Consequently, the position of the most advanced wetting front is not a direct measure of the volumetric degree of saturation. Local breakthrough may therefore occur well before the pore space approaches its limiting saturation, while substantial redistribution and air entrapment continue behind the leading edge of the wetting front. Consistently, within the resolved pore structure, the saturation approaches an apparent plateau, with the uptake rate becoming very small, rather than exhibiting a prolonged secondary imbibition regime. This suggests that the microstructural heterogeneity resolved here alone is insufficient to reproduce the anomalous long-term uptake behaviour observed experimentally \citep{hall1989sorptivity,alderete2020mechanism,zhangDifferentAnomaliesTwostage2024}.

The apparent limiting degree of saturation of approximately \(S\approx0.7\) should be interpreted within the physical assumptions of the present model rather than as a universal upper saturation limit for cement-based materials. The simulations directly demonstrate that a substantial fraction of the resolved pore volume becomes occupied by disconnected residual air. In cement-based materials, the capillary pore network is often assumed to become water-saturated at the end of primary imbibition, with residual air remaining primarily in larger voids \citep{smithDissolutionKineticsTrapped2020}. The present results do not necessarily contradict this picture, but show that substantial air entrapment can occur even within the capillary pore network during the early invasion stage. For the connected gas phase, the low-velocity conditions considered here imply negligible pressure-induced density variations, making the incompressible approximation well justified \citep{ferzigerComputationalMethodsFluid2020}. However, once gas becomes trapped, this assumption can affect the attainable saturation because its volume cannot decrease in response to increasing pressure. In real pore networks, gas compression could allow further liquid invasion early post-entrapment and could therefore lead to somewhat higher saturation already over the short timescales considered here. Moreover, dissolution may proceed rapidly for the smallest, highly pressurized gas volumes, so some contribution cannot be excluded even within the present simulation window. However, this would likely quickly increase the local concentration of dissolved gas and thereby reduce the driving force for further dissolution \citep{yu2023bubble}. Subsequent diffusive removal through the complex pore network may then become rate-limiting, so the continued reduction of residual gas volume and higher saturation are expected to occur over much longer timescales than considered in the present simulations \citep{smithDissolutionKineticsTrapped2020,joewondo2026coupled}. The present FIB--SEM domain also captures only a limited range of the pore-size spectrum, excluding both finer gel porosity and larger-scale features, so interactions across the broader multiscale pore structure of cement-based materials may further influence the attainable saturation. Possible moisture-induced microstructural evolution, including deformation and associated changes in pore geometry and connectivity, is also not considered \citep{alderete2024neutron,yin2024moisture}. The simulated value of \(S\approx0.7\) should therefore be regarded primarily as the saturation state resulting from early pore-scale invasion and trapping under the present assumptions.

This interpretation does not reduce the significance of the simulated saturation behaviour. Rather, the DNS identifies the physical origin of the initially trapped gas and separates the rapid capillary-invasion stage from slower processes that would subsequently modify the residual gas volume. The results therefore provide a mechanistic basis for distinguishing between the initial formation of trapped air and its subsequent evolution through rapid pressure-driven compression and slower dissolution and diffusive transport \citep{fagerlund2004service,governo2026revealing}.

Further differences arise between the idealized initial state considered here and moisture transport in real cementitious materials. The simulations assume an initially dry pore space, which provides a well-defined fundamental problem but is uncommon under most service conditions. Pre-existing moisture would alter the initial liquid connectivity, meniscus locations, and amount of gas that must be displaced during subsequent water ingress \citep{ren2023initial}, while thin liquid films along hydrophilic pore walls and partially saturated fine pores may provide additional pathways for liquid redistribution and facilitate local wetting ahead of the advancing menisci \citep{cai2022capillary,zhao2024dynamic}. In addition, the present simulations consider only bulk liquid-water transport. Vapour diffusion, adsorption and desorption, and capillary condensation within finer pore domains are not represented \citep{chidiac2025kinetic,wang2024understanding}. Although these processes may contribute to moisture redistribution in partially saturated cement-based materials, their influence is expected to remain secondary over the short timescales considered here.

\subsection{Computational considerations for pore-scale DNS}

Despite the relatively small physical volume of the simulated microstructure (approximately \(23.8\times9.7\times14.5~\mu\mathrm{m}^3\)), this level of physical and spatial detail comes at substantial computational cost. The principal three-dimensional pore-space domain considered in this study was discretized into approximately 38 million finite-volume cells. Three velocity components, pressure, volume fraction, and an additional capillary-pressure field, computed from a separate Poisson equation as part of the numerical treatment of surface tension \citep{raeini2012modelling,maes2021geochemfoam}, are solved, resulting in approximately 230 million algebraic unknowns per time step. The highly refined pore-scale mesh, required to explicitly resolve sub-micron capillary pores in the FIB-SEM geometry reconstructed at a voxel resolution of 29~nm, together with the relatively high surface tension of the air–water system, results in an average time step of approximately 0.1~ns to accurately and stably resolve interface advection and surface-tension-driven dynamics \citep{denner2015numerical}. As a result, reaching a physical simulation time of 40~\textmu s required approximately 400,000 time steps and about four weeks of computation on 128 CPU cores. The complete transient dataset generated for this simulation, including the fields required to reconstruct the animation provided in the Supplementary Information, amounted to approximately 1.6~TB. Such computational and data requirements necessarily limit the number of high-resolution three-dimensional virtual experiments that can be performed systematically. In contrast, two-dimensional domains and smaller three-dimensional geometries are substantially less expensive and therefore provide practical platforms for targeted mechanistic studies and systematic variations of geometry, scale, boundary conditions, fluid properties, and interfacial or material parameters. Consequently, despite the continuing increase in available computational resources, direct simulation of multiphase flow in realistic cementitious pore networks remains computationally demanding and is currently restricted to relatively small domains and short physical timescales. Further advances in high-performance computing will continue to push the practical limits of pore-scale DNS \citep{maes2026scalable,armstrong2026lattice}; however, the restrictive time-step requirements associated with resolving rapid capillary-driven interface dynamics at the pore scale remain a particularly severe computational bottleneck. More efficient and adaptive time-stepping strategies \citep{malenica2021full,maes2021geochemfoam,wang2026dynamic} will therefore be especially important for extending simulations towards larger domains and longer physical times. The principal role of pore-scale DNS should therefore not be to replace larger-scale transport models, but to identify the mechanisms and physical relationships that such models ultimately need to represent.

\subsection{Future opportunities for pore-scale DNS}

The most direct extensions of the present framework concern physical processes not represented in the current simulations, particularly those governing the evolution of trapped air after entrapment. Introducing gas compressibility would allow isolated air volumes to respond dynamically to the capillary and liquid-pressure fields already resolved by the present model, while coupling gas dissolution and diffusion throughout the liquid phase would permit the subsequent evolution of these trapped gas volumes over longer timescales. Together, these extensions would allow the effects of post-entrapment gas evolution on residual air volume and the attainable degree of saturation to be quantified directly.

Surface wettability represents another important physical factor to consider. The present simulations deliberately prescribe a uniform equilibrium contact angle of \(\theta=0^\circ\), providing the maximum capillary driving force and allowing the influence of pore geometry to be isolated. Real cementitious surfaces are chemically and mineralogically heterogeneous and may exhibit spatial variations in wettability together with contact-angle hysteresis \citep{kabir2023rapid,wang2023molecular,behnoudfar2022contact}. Within the cementitious pore space, these variations may influence contact-line motion and local invasion pathways, although the corresponding range of equilibrium contact angles within a single cementitious microstructure is expected to remain relatively limited. Larger wettability contrasts may arise at the steel--concrete interface, where the steel surface can exhibit substantially different wetting behaviour from the surrounding cementitious matrix, thereby altering the local moisture distribution \citep{malenicaDirectNumericalModelling2024,governo2026revealing}.

The resulting local moisture heterogeneity is particularly relevant to corrosion and also illustrates how the DNS framework could be extended beyond moisture transport alone. The simulations show that neighbouring regions of the reinforcing steel may simultaneously be water-covered or remain adjacent to air-filled pores, meaning that a single bulk degree of saturation cannot uniquely describe the local moisture environment relevant to corrosion processes. Coupling the present framework with heterogeneous wettability, oxygen and ionic-species transport and electrochemical reactions would enable direct investigation of how pore-scale moisture redistribution controls local corrosion conditions \citep{schmid2026pore}.

A further extension would be to consider partially saturated initial states and additional moisture-transport mechanisms beyond liquid-water imbibition. Including vapour transport and, at finer scales, adsorption, desorption, and capillary condensation would broaden the range of moisture conditions accessible to pore-scale DNS.

The directly resolved pressure fields additionally suggest opportunities for coupling moisture transport with mechanical processes. The simulations resolve both capillary-induced liquid underpressures and overpressures within trapped gas volumes. These local fluid pressures act on the surrounding solid skeleton and could, in future, be coupled to mechanical descriptions of hydration products and pore walls to investigate hydro-mechanical interactions at the pore scale \citep{zhou2019multiscale,zou2021capillary,gao2022incorporating}.

As fluid--solid interactions progress, the assumption of a static pore geometry may eventually become limiting. Cement-based materials can undergo continued hydration, dissolution and precipitation, shrinkage, damage, and other microstructural changes that modify pore connectivity, local wettability, and transport pathways. Accounting for such evolving pore geometries would require substantially more complex coupling between multiphase flow, chemical reactions, and solid-phase evolution \citep{molins2021simulation,patel2021multi,soulaine2024microcontinuum}, but could ultimately provide a route towards describing moisture transport together with the progressive physicochemical evolution of cementitious microstructures.

Overall, the principal contribution of the present work is not simply the numerical reproduction of spontaneous imbibition within a realistic pore geometry, but the ability to expose and separate the pore-scale mechanisms from which the observed transport behaviour emerges. Image-based DNS reveals how irregular pore geometry drives continuous interface reconfiguration, how interface curvature generates strongly heterogeneous capillary-pressure fields, how the governing force balance changes with scale, how three-dimensional connectivity modifies local capillary barriers, and how multiple gas-trapping pathways progressively produce residual air entrapment. At the same time, the simulations identify which additional physical processes may become important when moving beyond the conditions considered in the present study. In this sense, pore-scale DNS provides a framework for linking fundamental capillary physics and experimentally reconstructed microstructure to physically based descriptions of moisture transport and durability-relevant processes in cement-based materials.

%% file: 05_Conclusions.tex
\section{Conclusions}
\label{sec_conclusions}

This study used image-based pore-scale Navier--Stokes/VOF simulations to investigate spontaneous air--water imbibition within the capillary porosity of a realistic three-dimensional FIB--SEM-resolved cementitious pore structure at nanometre-to-micrometre pore scales and sub-nanosecond time stepping. To the best of our knowledge, this provides the first direct view of the corresponding transient transport dynamics at such spatio-temporal scales, which remain inaccessible to experimental observation and have not previously been captured by numerical modelling.

The results show that capillary imbibition in irregular pore networks cannot be described simply in terms of characteristic pore size. This departs from the simplified picture underlying classical analytical capillary models, in which pore geometry constrains meniscus curvature and capillary pressure. Instead, as the interface advances through the irregular pore geometry, it continuously reconfigures, generating transient and spatially heterogeneous curvature and capillary-pressure fields that govern local interface motion and liquid redistribution.

The simplified two-dimensional pore-scale analyses further show that local imbibition dynamics arise from a scale-dependent competition between capillary, viscous, inertial, and gravitational effects. Their relative importance changes markedly with pore scale, leading to different limiting mechanisms across the range considered. The corresponding three-dimensional simulations further demonstrate that the manifestation of these local mechanisms can be substantially modified by the full pore-network geometry. Local barriers that arrest the isolated two-dimensional meniscus do not necessarily prevent continued wetting-front advancement in the full pore network. Three-dimensional interface curvature, the spatially extended nature of three-dimensional contact lines, and alternative connected pathways allow the wetting front to reconfigure and redistribute around locally unfavourable configurations. The two-dimensional analyses are therefore particularly valuable for isolating and understanding individual pore-scale mechanisms, whereas the three-dimensional simulations reveal how these mechanisms collectively govern imbibition within realistic pore networks.

Air entrapment emerges as an inherent consequence of the evolving network dynamics and is not restricted to particular pore types. Gas becomes isolated through bypass and reconnection around slowly filling regions, typically associated with locally larger pores, sealing of dead-end pores, or more generally through interface reconfiguration and coalescence that disconnect the gas phase in a variety of pore configurations. As imbibition proceeds, preferential invasion of the most favourable connected pathways and progressive gas isolation increasingly constrain further liquid invasion, contributing to the declining saturation rate. Importantly, substantial residual air persists already at the scale of capillary pores, even under perfectly water-wetting conditions.

Once isolated, trapped-gas pressure is governed by the surrounding liquid pressure and the curvature of the confining air--water interface. Analysis of 156 trapped-air clusters shows that capillary pressure cannot be reliably predicted from pore size or trapped-air volume alone: only 24 fall within \(\pm10\%\) of the equivalent-volume spherical-bubble prediction. This discrepancy arises because the surrounding solid geometry can impose strongly curved air--water interfaces, as in dead-end pores, whereas detached, nearly spherical bubbles of comparable volume can exhibit substantially lower curvature and capillary pressure. Consequently, capillary pressure can reach several megapascals at micrometre and sub-micrometre scales and differ substantially even between trapped-air clusters of nearly identical volume. 
However, the trapped-gas volumes, associated pressures, and apparent limiting saturation reached within the simulated domain reflect the initial rapid capillary-invasion and trapping stage under the incompressible-gas assumption and within the spatial and temporal scales considered here, and should not be interpreted as the ultimate trapped-air state or a general saturation limit for cement-based materials.

The steel--concrete interface analysis further demonstrates how the same framework can be used to resolve local moisture conditions directly at durability-relevant interfaces. By identifying the water-covered steel surface and the surrounding air--water configuration, the simulations provide a pore-scale description of the potentially electrochemically active steel area and its proximity to neighbouring air-filled regions that may provide local access to gaseous oxygen.

Overall, the study demonstrates how image-based pore-resolved DNS can serve as a virtual experiment for linking experimentally reconstructed microstructure with the fundamental physics of capillary imbibition. By resolving interface geometry and evolution, flow and pressure fields, phase connectivity, and gas entrapment simultaneously, the approach provides direct access to pore-scale mechanisms that remain difficult to isolate experimentally and offers a basis for interpreting moisture transport and local durability-relevant conditions in cement-based materials. It also provides a foundation for future extensions towards increasingly complex coupled processes, including reactive transport, electrochemical reactions, and other multiphysics phenomena relevant to durability.